\documentclass[10pt, conference]{IEEEtran}
\IEEEoverridecommandlockouts

\usepackage{cite}
\usepackage{amsmath,amssymb,amsfonts}
\usepackage{algorithm}
\usepackage{algorithmic}
\usepackage{graphicx}
\usepackage{textcomp}
\usepackage{xcolor}
\usepackage{listings}
\usepackage{float}
\usepackage{subcaption}
\usepackage{booktabs}
\usepackage{makecell}
\usepackage{multirow}
\usepackage{tikz}
\usepackage{url}
\usetikzlibrary{arrows.meta, positioning, tikzmark}

\newif\ifanonymous
\def\BibTeX{{\rm B\kern-.05em{\sc i\kern-.025em b}\kern-.08em
    T\kern-.1667em\lower.7ex\hbox{E}\kern-.125emX}}

\begin{document}

\title{Prompt Less, Infer More: Static Analysis-Guided LLM Type Annotation for Python}

\ifanonymous
\author{
\IEEEauthorblockN{Anonymous Authors}
}
\else
\author{
    
\IEEEauthorblockN{Wonseok Oh}
\IEEEauthorblockA{Korea University \\
Seoul, Republic of Korea \\
wonseok\_oh@korea.ac.kr}

\and

\IEEEauthorblockN{Michael Pradel}
\IEEEauthorblockA{CISPA Helmholtz Center for Information Security \\
Stuttgart, Germany \\
michael@binaervarianz.de}

\and

\IEEEauthorblockN{Hakjoo Oh\textsuperscript{*}}
\IEEEauthorblockA{Korea University \\
Seoul, Republic of Korea \\
hakjoo\_oh@korea.ac.kr}
\thanks{\textsuperscript{*}Corresponding author}

}
\fi

\newcommand{\myparagraph}[1]{\par\smallskip\noindent\textit{\textbf{#1.}}}
\newcounter{example}
\newenvironment{example}
  {\refstepcounter{example}\par\smallskip\noindent\textbf{Example~\theexample.}\ }
  {\par\smallskip}

\newcommand{\tool}{\textsc{Salt}}
\newcommand{\ours}{\tool{}}
\newcommand{\python}{Python}
\newcommand{\llm}{LLM}
\newcommand{\llms}{LLMs}

\newcommand{\tiger}{\textsc{Tiger}}
\newcommand{\typegen}{\textsc{TypeGen}}
\newcommand{\typetfive}{\textsc{TypeT5}}
\newcommand{\typetfivegpt}{\textsc{TypeT5-GPT}}
\newcommand{\hityper}{\textsc{HiTyper}}
\newcommand{\nester}{\textsc{Nester}}
\newcommand{\aider}{\textsc{Aider}}
\newcommand{\claudecode}{\textsc{Claude Code}}
\newcommand{\codexcli}{\textsc{Codex CLI}}
\newcommand{\githubcopilot}{\textsc{GitHub Copilot}}
\newcommand{\cursor}{\textsc{Cursor}}
\newcommand{\pyrefly}{\textsc{Pyrefly}}
\newcommand{\jedi}{\textsc{Jedi}}
\newcommand{\typybench}{\textsc{TypyBench}}

\newcommand{\exactacc}{Exact Accuracy}
\newcommand{\baseacc}{Base Accuracy}
\newcommand{\todo}[1]{\textcolor{red}{[TODO: #1]}}

\newif\ifshowmodifications
\showmodificationsfalse
\definecolor{modificationcolor}{RGB}{166,57,16}
\newcommand{\modify}[1]{%
  \ifshowmodifications
    \par\smallskip
    {\color{modificationcolor}\small\noindent\textbf{[MODIFY]} #1\par}%
    \smallskip
  \fi
}

\definecolor{revisioncolor}{RGB}{145,30,105}
\newcommand{\revise}[2]{%
  \ifshowmodifications
    {\color{revisioncolor}{\small\textbf{[REVISED: #1]}} #2}%
  \else
    #2%
  \fi
}

\newcommand{\revisenote}[2]{%
  \ifshowmodifications
    \revise{#1}{#2}%
  \fi
}

\definecolor{codebg}{RGB}{248,249,251}
\definecolor{codeframe}{RGB}{214,220,229}
\definecolor{codenumber}{RGB}{118,128,140}
\definecolor{codekeyword}{RGB}{34,92,171}
\definecolor{codestring}{RGB}{151,80,20}
\definecolor{codecomment}{RGB}{88,129,87}

\lstdefinestyle{pythoncode}{
    language=Python,
    morekeywords={bytes},
    basicstyle=\ttfamily\scriptsize\linespread{0.85}\selectfont,
    keywordstyle=\color{codekeyword}\bfseries,
    stringstyle=\color{codestring},
    commentstyle=\color{codecomment}\itshape,
    columns=fullflexible,
    keepspaces=true,
    showstringspaces=false,
    numbers=left,
    numberstyle=\tiny\color{codenumber},
    numbersep=0.55em,
    stepnumber=1,
    backgroundcolor=\color{codebg},
    frame=single,
    rulecolor=\color{codeframe},
    framerule=0.35pt,
    framesep=3pt,
    framexleftmargin=1.35em,
    framexrightmargin=0.35em,
    xleftmargin=\dimexpr0.015\columnwidth+1.35em\relax,
    xrightmargin=0pt,
    linewidth=0.97\columnwidth,
    aboveskip=0.6em,
    belowskip=0.6em,
    breaklines=true,
    breakatwhitespace=true,
    tabsize=4
}

\lstdefinestyle{pythoninline}{
    language=Python,
    morekeywords={bytes},
    basicstyle=\ttfamily\small,
    keywordstyle=\color{codekeyword}\bfseries,
    stringstyle=\color{codestring},
    commentstyle=\color{codecomment}\itshape,
    columns=fullflexible,
    keepspaces=true,
    showstringspaces=false
}
\lstMakeShortInline[style=pythoninline]|

\captionsetup[figure]{labelsep=period}
\captionsetup[subfigure]{labelformat=parens,labelsep=space,skip=2pt}

\newcommand{\seq}[1]{\ensuremath{\overrightarrow{#1}}}
\newcommand{\Unknown}{\ensuremath{\mathsf{Unknown}}}
\newcommand{\Var}{\ensuremath{\textsl{Var}}}
\newcommand{\NTProgram}{\ensuremath{P}}
\newcommand{\NTClass}{\ensuremath{K}}
\newcommand{\NTMethod}{\ensuremath{M}}
\newcommand{\NTExpr}{\ensuremath{E}}
\newcommand{\NTStmt}{\ensuremath{S}}
\newcommand{\Classes}[1]{\ensuremath{\mathsf{classes}(#1)}}
\newcommand{\Methods}[1]{\ensuremath{\mathsf{methods}(#1)}}
\newcommand{\ParamSlots}[1]{\ensuremath{\mathsf{params}(#1)}}
\newcommand{\ReturnSlots}[1]{\ensuremath{\mathsf{returns}(#1)}}
\newcommand{\Slots}[1]{\ensuremath{\mathsf{slots}(#1)}}
\newcommand{\Annots}[1]{\ensuremath{\mathsf{annots}(#1)}}
\newcommand{\Types}{\ensuremath{\mathcal{T}}}
\newcommand{\FHint}{\ensuremath{F}}
\newcommand{\Hint}{\FHint}
\newcommand{\BStrong}{\ensuremath{B_s}}
\newcommand{\BWeak}{\ensuremath{B_w}}
\newcommand{\Ann}{\ensuremath{A}}
\newcommand{\ReturnSlot}[1]{\ensuremath{\mathsf{ret}(#1)}}
\newcommand{\Cand}{\ensuremath{\mathsf{Cand}}}
\newcommand{\LitVal}{\ensuremath{\ell}}
\newcommand{\CallExpr}{\ensuremath{\kappa}}
\newcommand{\GetSlots}{\ensuremath{\mathsf{get\_slots}}}
\newcommand{\Lit}{\ensuremath{\mathsf{lit}}}
\newcommand{\IsInstance}{\ensuremath{\mathsf{isinstance}}}
\newcommand{\TypeJudgment}[3]{\ensuremath{#1 \vdash #2 \Rightarrow #3}}
\newcommand{\ForwardExpr}[5]{\ensuremath{#1;#2 \vdash #3 \Rightarrow #4;#5}}
\newcommand{\StrongEvidence}[2]{\ensuremath{#1 \leadsto_s #2}}
\newcommand{\WeakEvidence}[2]{\ensuremath{#1 \leadsto_w #2}}
\newcommand{\HintUpdate}[2]{\ensuremath{\FHint\sqcup[#1\mapsto #2]}}
\newcommand{\AnnUpdate}[2]{\ensuremath{\Ann\sqcup[#1\mapsto #2]}}

\newcommand{\Func}{M}
\newcommand{\FuncName}{m}
\newcommand{\Param}{p}
\newcommand{\ParamSeq}{\seq{\Param}}

\maketitle

\begin{abstract}
Type annotations improve the maintainability and reliability of \python{} programs, but many real-world projects remain under-annotated because manual annotation is time-consuming and error-prone.
Recent neural and \llm{}-based type inference systems can recover missing annotations, yet they often trade accuracy for efficiency.
Existing approaches often fail to exploit project-wide type evidence, or do so inefficiently through unnecessary model calls and noisy prompts.
We present \tool{}, an \llm{}-based type inference system guided by static analysis.
\tool{} integrates static analysis with \llm{} inference to resolve types statically, prioritize \llm{} queries, and construct concise type-aware context.
To demonstrate the effectiveness of \tool{}, we evaluated it on 12 real-world \python{} projects.
\tool{} achieves the best accuracy among the evaluated systems and reduces estimated \llm{} cost by up to 78.5\% compared with a strong coding-assistant baseline.
It also shows similar cost savings with the frontier model GPT-5.5, demonstrating that \tool{} remains effective with strong \llm{}s.
\end{abstract}


\section{Introduction}
\label{sec:introduction}

Type annotations have become an important part of modern \python{} development. 
Although \python{} is dynamically typed, type annotations allow developers to catch bugs with static checkers and make large codebases easier to maintain \cite{xie2025type}. 
Unfortunately, many \python{} projects remain under-annotated~\cite{di2022evolution},
and manually annotating them is time-consuming and error-prone \cite{ore2018assessing}.
This problem has motivated a growing interest in automated type annotation for \python{}~\cite{lambdanet,deeplearning,class-dlinfer,class-typewriter,hityper,similar-type4py,similar-typlius,class-typebert,typet5,tiger,nester,typegen,pyrefly,rule-maxsmt,rule-starkiller,rule-pyinder,pyright,mypy,pytype}.

\myparagraph{Existing Approaches}
Recent work has explored specialized type inference tools based on neural~\cite{lambdanet,deeplearning,class-dlinfer,class-typewriter,hityper,similar-type4py,similar-typlius,class-typebert,tiger,typet5} and \llm{}-based~\cite{nester,typegen} approaches, which show promise in annotating real-world \python{} code.
More recently, general-purpose \llm{} coding assistants, such as \aider{}~\cite{aider}, \claudecode{}~\cite{claudecode}, \codexcli{}~\cite{codexcli}, \githubcopilot{}~\cite{githubcopilot}, and \cursor{}~\cite{cursor}, have become a natural alternative.
Although these tools are not designed specifically for type inference, type annotation is also a code-editing task: a tool that can inspect a repository and modify code can be prompted to add missing annotations.

We observed that these assistants can achieve strong accuracy on a benchmark of real-world \python{} projects.
As a preview of our results, the strongest existing specialized tool, \typetfive~\cite{typet5}, reaches 69.30\% accuracy, while \aider{}, an open-source coding assistant, reaches up to 84.13\% accuracy.
This result arises because \aider{} can retrieve and reason about broad repository context, including cross-file evidence, 
whereas specialized systems tend to focus only on the target function or fail to make use of diverse contextual information.

However, this accuracy comes at a high cost.
For a project with about 14k lines of code, \aider{} with GLM-4.7~\cite{glm47} consumed 24.83M input and 0.65M output tokens, costing approximately \$57.67\footnote{We calculate the cost based on the GLM-4.7 API pricing (https://www.cerebras.ai/pricing). Input: \$2.25/M, Output: \$2.75/M.}.
First, \aider{} invokes the \llm{} for every function, even trivial ones such as |def foo(): return 0|.
Second, \aider{} constructs context through lexical name matching rather than program analysis, producing verbose prompts that often include irrelevant information.

In summary, specialized tools are limited because they fail to capture relevant context,
whereas general-purpose coding assistants are inefficient because they trigger unnecessary \llm{} invocations
and flood the prompt with irrelevant information.

\myparagraph{This Work}
We present \tool{}, a system that uses \textbf{\underline{S}}tatic \textbf{\underline{A}}nalysis to guide \textbf{\underline{L}}LM-based \textbf{\underline{T}}ype annotation.
\tool{} retains the broad repository-level reasoning of general-purpose coding assistants, but uses static analysis to reduce unnecessary \llm{} usage.
\tool{} first performs project-wide forward and backward type inference.
Forward inference propagates caller-side evidence, while backward inference recovers use-site evidence that flows from callees.
When static evidence is sufficiently precise, \tool{} directly annotates the code without invoking an \llm{}.
For unresolved functions, \tool{} uses the inferred type information to prioritize high-leverage \llm{} queries and to construct concise type-aware context for each target function.

Our evaluation on 12 real-world \python{} projects from \typybench{}~\cite{typybench} shows that this design improves both effectiveness and efficiency.
We evaluate \tool{} with three \llm{} models of different sizes: Gemma 4~\cite{gemma4}, gpt-oss~\cite{gptoss}, and GLM-4.7~\cite{glm47}.
Across three \llm{} models, \tool{} improves accuracy over \aider{} by 3.5\% on average while reducing \llm{} calls by 35.7\%.
In addition, \tool{} reduces estimated cost by 75.7\% on average.
\tool{} also remains effective with the frontier model GPT-5.5, reducing estimated cost by 74.8\% while slightly improving accuracy over \aider{} (also backed by GPT-5.5).
\revise{B3}{Moreover, our ablation study shows that static analysis, query prioritization, and type-aware context construction each improve accuracy and reduce \llm{} usage.
}

\myparagraph{Contributions}
This paper makes the following contributions:
\begin{itemize}
  \item We show that general-purpose coding assistants can outperform specialized type inference tools, but at substantial cost from unnecessary \llm{} calls and verbose context.
  \item We present \tool{}, which uses static analysis to actively guide \llm{}-based annotation through (i) forward and backward inference, (ii) query prioritization, and (iii) concise type-aware context.
  \item We evaluate \tool{} on 12 real-world \python{} projects with four \llm{}s, showing improved accuracy over strong baselines with up to 78.5\% lower cost, including with the frontier model GPT-5.5.
\end{itemize}

\section{Overview of \tool{}}
\label{sec:overview}

\tool{} addresses the problem of annotating untyped or partially typed \python{} functions.
We focus on function-level annotations rather than variable annotations because prior work has shown that developers mostly annotate functions~\cite{di2022evolution}.
For example, given an unannotated function, the goal is to infer its missing parameter and return annotations:
\begin{center}
\begin{minipage}{0.42\linewidth}
\centering
\textit{Unannotated}
\begin{lstlisting}[style=pythoncode,numbers=none,linewidth=\linewidth,xleftmargin=0pt,framexleftmargin=0.35em]
from pathlib import Path
def foo(p):
  return Path(p+".py")
\end{lstlisting}
\end{minipage}
\quad
$\Rightarrow$
\quad
\begin{minipage}{0.42\linewidth}
\centering
\textit{Annotated}
\begin{lstlisting}[style=pythoncode,numbers=none,linewidth=\linewidth,xleftmargin=0pt,framexleftmargin=0.35em]
from pathlib import Path
def foo(p: str) -> Path:
  return Path(p+".py")
\end{lstlisting}
\end{minipage}
\end{center}
\tool{} aims to make this process both effective and efficient: it should infer
accurate annotations by using project-wide evidence, while minimizing unnecessary \llm{} usage.
\tool{} achieves this by combining static type analysis with \llm{}-based annotation in a synergistic way.

\subsection{Motivating Example}

Prior work has recognized that type prediction benefits from program context beyond the target function.
Existing approaches therefore use techniques such as forward type inference for pre-annotation, caller-callee context construction, and callee-first query ordering.
However, we found that these techniques remain suboptimal in how they use type information.
First, forward type inference captures only caller-to-callee evidence and misses useful evidence that flows backward.
Second, existing approaches do not sufficiently use type analysis results when deciding what context to include and which function to query next.
Figure~\ref{fig:motivation-example} illustrates these limitations.

\begin{figure}[t]
\centering
\begin{lstlisting}[style=pythoncode]
class String: def kind(self, name: str)   -> str: ...
class Bytes:  def kind(self, name: bytes) -> bytes: ...
def get_paths() -> list[str]: ...

def internal(name):
  x = get_paths()
  paths = pack(x, name, "i-str")
  return paths

def external(name): 
  x = external_func()
  paths = pack(x, name, "e-str")
  return paths

def pack(paths, name, msg): 
  print_log(name, msg)
  k = String()
  return [(norm(p), k.kind(name)) for p in paths]

def norm(path): 
  return path.rstrip()

def print_log(name, msg) -> None: 
  print(f"{name}: {msg}")
\end{lstlisting}
\caption{A motivating example where existing approaches either miss useful static evidence or spend unnecessary \llm{} calls.
}
\vspace{-1em}
\label{fig:motivation-example}
\end{figure}

\begin{figure*}
\centering
\begin{tikzpicture}[
  font=\scriptsize,
  stage/.style={
    draw=codeframe,
    fill=codebg,
    thick,
    minimum width=2.35cm,
    minimum height=0.72cm,
    align=center
  },
  arrow/.style={-Stealth, thick}
]
\node[stage] (input) {Input\\Python Project};
\node[stage, right=0.35cm of input] (analysis) {\textbf{(1)} Forward \&\\Backward Analysis};
\node[stage, right=0.35cm of analysis] (rank) {\textbf{(2)} Prioritize\\LLM Targets};
\node[stage, right=0.35cm of rank] (context) {\textbf{(3)} Type-Guided\\Context};
\node[stage, right=0.35cm of context] (llm) {LLM\\Annotation};
\node[stage, right=0.35cm of llm] (output) {Annotated\\Project};

\draw[arrow] (input) -- (analysis);
\draw[arrow] (analysis) -- (rank);
\draw[arrow] (rank) -- (context);
\draw[arrow] (context) -- (llm);
\draw[arrow] (llm) -- (output);
\draw[arrow] (llm.south) -- ++(0,-0.45cm) -| (analysis.south);
\end{tikzpicture}
\caption{Workflow of \tool{}. The system repeatedly alternates between static type analysis and \llm{} annotation until the remaining annotations are resolved.}
\vspace{-1em}
\label{fig:workflow}
\end{figure*}
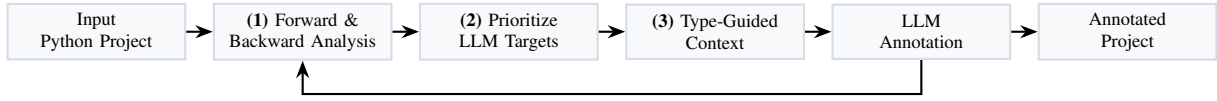

\begin{figure}[t]
\centering
\newcommand{\namehl}[1]{\begingroup\setlength{\fboxsep}{0.0pt}\colorbox{green!12}{\strut #1}\endgroup}
\newcommand{\msghl}[1]{\begingroup\setlength{\fboxsep}{0.0pt}\colorbox{cyan!12}{\strut #1}\endgroup}
\begin{lstlisting}[style=pythoncode,escapeinside={(*@}{@*)}]
class String: def kind(self, name: str)   -> str: ...
class Bytes:  def kind(self, name: bytes) -> bytes: ...
def get_paths() -> list[str]: ...

def internal((*@\namehl{\tikzmarknode{intName}{\texttt{name}}\texttt{: str}}@*)):
  x = get_paths()
  paths = pack(x, name, (*@\tikzmarknode{litI}{}@*)"i-str")
  return paths

def external((*@\namehl{\tikzmarknode{extName}{\texttt{name}}\texttt{: str}}@*)):
  x = external_func()
  paths = pack(x, name, "e-str"(*@\tikzmarknode{litE}{}@*))
  return paths

def pack(paths, (*@\namehl{\tikzmarknode{packName}{\texttt{name}}\texttt{: str}}@*), (*@\msghl{\tikzmarknode{packMsg}{\texttt{msg}}\texttt{: str}}@*)):
  print_log(name, msg)
  k = String()
  return [(norm(p), k.(*@\tikzmarknode{kindCall}{kind(name)}@*)) for p in paths]

def norm(path):
  return path.rstrip()

def print_log((*@\namehl{\tikzmarknode{logName}{\texttt{name}}\texttt{: str}}@*), (*@\msghl{\tikzmarknode{logMsg}{\texttt{msg}}\texttt{: str}}@*)) -> None:
  print(f"{name}: {msg}")
\end{lstlisting}
\begin{tikzpicture}[
  remember picture,
  overlay,
  fwd/.style={-Stealth, thick, color=black, shorten <=1.5pt, shorten >=1.5pt},
  bwd/.style={-Stealth, very thick, dash pattern=on 2.5pt off 1.2pt, color=black, shorten <=1.5pt, shorten >=1.5pt},
  note/.style={font=\scriptsize, fill=white, inner sep=1pt}
]
\draw[fwd] (litI.south) to[bend right=22] (packMsg.north);
\draw[fwd] (litE.south) to[bend right=16] (packMsg.north);
\draw[fwd] (packMsg.south) to[bend left=18] (logMsg.north);

\draw[bwd] (kindCall.north) to[bend left=20] (packName.south);
\draw[fwd] (packName.south) to[bend right=18] (logName.north);
\draw[bwd] (packName.north) to[bend right=22] (intName.south);
\draw[bwd] (packName.north) to[bend left=22] (extName.south);
\end{tikzpicture}
\vspace{-1em}
\caption[Type propagation over the motivating example.]{Type propagation over the motivating example. \protect\tikz[baseline=-0.5ex]{\draw[-Stealth, thick] (0,0) -- (0.55,0);} denotes forward propagation, and \protect\tikz[baseline=-0.5ex]{\draw[-Stealth, very thick, dash pattern=on 2.5pt off 1.2pt] (0,0) -- (0.55,0);} denotes backward propagation.}
\label{fig:motivation-propagation}
\end{figure}

\myparagraph{Limitation of Forward Type Inference}
We found that forward type inference is insufficient to capture all the relevant evidence in real-world code.
For example, \hityper{} \cite{hityper} uses forward type inference to
pre-annotate types that flow from callers to callees,
such as the parameter |msg| of |pack| at line~15, which is passed string literals at lines~7 and~12.
Subsequently, their approach propagates this annotation to the callee |print_log| at line~16, yielding the parameter annotation |msg: str| in |print_log|.

However, since |pack| constructs |k = String()| and calls |k.kind(name)|, the
second parameter |name| of |pack| should be |str|.
This evidence comes from the resolved callee |String.kind| and can be propagated backward to |pack|,
but a forward-only approach cannot leverage this information.

\myparagraph{Limitation of LLM Querying}
Existing approaches also fail to maximize the synergy between \llm{} queries and type analysis.

The first issue is query ordering.
Prior work such as \typetfive{} \cite{typet5} considers a callee-first order,
where the \llm{} is queried for leaf functions before their callers.
In Figure~\ref{fig:motivation-example}, this strategy would query |norm| at line~20 before |pack|.
However, annotating |norm| first has limited leverage: 
even if the model correctly infers |norm: str -> str|, 
this annotation does not determine the type of |paths| in |pack|, so another \llm{} call is still needed for |pack|.

The second issue is context construction.
Existing approaches often limit the prompt to directly related caller and callee functions due to prompt size constraints.
Suppose an existing approach queries |norm| first.
The caller-callee context includes |pack| as the caller and no callees.
However, this context still does not reveal the type of parameter |paths| precisely,
because |pack| receives |paths| from its callers, such as |internal| and |external|.
Without type analysis results that summarize these caller-side facts, the model
has limited evidence for inferring the parameter type of |norm|.

Even when querying |pack|, existing context construction is inefficient.
\typetfive{} \cite{typet5} and \aider{} \cite{aider} construct caller-callee context through lexical name matching instead of type analysis.
For the call |k.kind(name)| in |pack|, they retrieve all functions named |kind|, including both |String.kind| and the irrelevant |Bytes.kind|.
However, only |String.kind| is relevant because |k| is constructed as |String| at line~17, and including |Bytes.kind| introduces noise into the prompt.

\subsection{Illustration of Key Ideas}\label{sec:overview-key-ideas}
To address these limitations, \tool{} combines static analysis with \llm{} querying in three ways (Figure~\ref{fig:workflow}):
\textbf{(1)} \tool{} first extends pre-annotation beyond forward propagation by exploiting backward evidence from callees.
\textbf{(2)} It then uses the inferred type information to select high-leverage target functions that are worth querying the \llm{} for.
\textbf{(3)} Finally, \tool{} uses the inferred type information to construct type-aware context before querying the model.
We illustrate these ideas using the motivating example below.

\begin{figure}[t]
\centering
\newcommand{\llmhl}[1]{\begingroup\setlength{\fboxsep}{0.0pt}\colorbox{yellow!18}{\strut #1}\endgroup}
\newcommand{\statichl}[1]{\begingroup\setlength{\fboxsep}{0.0pt}\colorbox{green!12}{\strut #1}\endgroup}
\begin{lstlisting}[style=pythoncode,escapeinside={(*@}{@*)}]
def internal(name: str) (*@\statichl{\tikzmarknode{intRet}{\texttt{-> list[tuple[str, str]]}}}@*):
  ...
  paths = pack(x, name, "i-str")
  return paths

def external(name: str) (*@\statichl{\tikzmarknode{extRet}{\texttt{-> list[tuple[str, str]]}}}@*):
  ...
  paths = pack(x, name, "e-str")
  return paths

def pack((*@\llmhl{\tikzmarknode{packPaths}{\texttt{paths}}\texttt{: list[str]}}@*), name: str, msg: str)
                          (*@\llmhl{\tikzmarknode{packRet}{\texttt{-> list[tuple[str, str]]}}}@*):
  ...
  return [(norm(p), k.kind(name)) for p in paths]

def norm((*@\statichl{\tikzmarknode{normPath}{\texttt{path}}\texttt{: str}}@*)) (*@\statichl{\tikzmarknode{normRet}{\texttt{-> str}}}@*): ...
\end{lstlisting}
\begin{tikzpicture}[
  remember picture,
  overlay,
  prop/.style={-Stealth, thick, color=black, shorten <=1.5pt, shorten >=1.5pt}
]
\draw[prop] (packPaths.south) to[bend left=20] (normPath.north);
\draw[prop] (packRet.north) to[bend right=20] (intRet.south);
\draw[prop] (packRet.north) to[bend left=20] (extRet.south);
\end{tikzpicture}
\vspace{-1em}
\caption{Static propagation after querying \texttt{pack}.}
\vspace{-1em}
\label{fig:pack-propagation}
\end{figure}

\myparagraph{(1) Combining Forward and Backward Inference}
By incorporating backward analysis, \tool{} can recover evidence that forward
propagation alone misses in Figure~\ref{fig:motivation-example}.
Figure~\ref{fig:motivation-propagation} summarizes how these annotations are
propagated.

Forward analysis infers |msg: str| for |pack| and |print_log| from string literals passed by |internal| and |external|.
Backward analysis then uses the resolved callee |k.kind(name)| at line~18 to infer |name: str| for |pack| and propagates it back to |internal| and |external|.
Moreover, a subsequent round of forward propagation uses the inferred |name: str| to annotate parameter |name| of |print_log| at line~23. 
Thus, |print_log| becomes fully annotated without an additional \llm{} query.

A key observation is that backward information should be interpreted together with forward information.
For example, consider the following code:

\begin{lstlisting}[style=pythoncode,caption={Combining Forward and Backward Evidence},label={lst:combined-evidence},belowskip=0.4em]
def foo(path, encoding):
  if isinstance(encoding, bytes):
    encoding = encoding.decode()
  open(path, encoding=encoding)

foo("src/a.py", "utf-8")
\end{lstlisting}

The function |open| at line~4 takes a path-like object (|str|, |bytes|, or |PathLike|) for its first argument.
If we propagate the type of parameter |path| backward only from |open|, we infer a broad type such as |str| $\mid$ |bytes| $\mid$ |PathLike|.
However, such a broadly inferred type is not useful in practice.
Combining this backward evidence with caller-side evidence avoids this over-approximation.
Since |foo| is called with the string literal |"src/a.py"|, 
the type of |path| can be narrowed more precisely to |str|.
Conversely, caller-side evidence alone suggests |encoding: str| because |foo| is called with |"utf-8"| at line~6.
However, the function body shows that |encoding| can also be |bytes| through the |isinstance(encoding, bytes)| check.
Combining these two sources of evidence allows \tool{} to infer the more appropriate type |str| $\mid$ |bytes| for |encoding|.

\begin{figure}[t]
\centering
\begin{lstlisting}[style=pythoncode]
# Type hints from static analysis:
# paths <- list[str]  # from internal
# paths <- Unknown    # from external

class String:
  def kind(self, name: str) -> str: ...

def internal(name: str):
  x = get_paths()
  paths = pack(x, name, "i-str")
  return paths

def external(name: str):
  x = external_func()
  paths = pack(x, name, "e-str")
  return paths

def pack(paths, name: str, msg: str):
  print_log(name, msg)
  k = String()
  return [(norm(p), k.kind(name)) for p in paths]

def norm(path):
  return path.rstrip()

def print_log(name, msg) -> None: ...
\end{lstlisting}
\caption{Type-guided context construction for querying \texttt{pack}. \tool{} includes the type-resolved callee \texttt{String.kind} and concise static hints, without adding irrelevant callees or extra caller-side function bodies.}
\vspace{-0.5em}
\label{fig:context-ours}
\end{figure}

\myparagraph{(2) Prioritizing LLM Calls to Maximize Static Inference}
To maximize the leverage of static analysis,
we determine the order of \llm{} calls based on how many other functions each function can influence.

When selecting a function for an \llm{} query, \tool{} considers how many other functions can benefit from its annotation.
Suppose that \tool{} selects the next \llm{} target after the static propagation shown in Figure~\ref{fig:motivation-propagation}.
In this state, |pack| is a high-impact target because its annotation affects
both its callers, |internal| and |external|, and its callee, |norm|.
\tool{} therefore queries the \llm{} for |pack| first, asking the model to infer
the type of |paths| and the return type of |pack|.
If this query succeeds (|paths: list[str]| and \texttt{return:} |list[tuple[str, str]]|), subsequent static propagation uses
|paths: list[str]| to infer that the first argument of |norm| at line~20 is |str| from the list comprehension at line~18.
It also propagates the return type of |pack| to lines~7 and~12, thereby
statically annotating the return types of |internal| and |external|.
Figure~\ref{fig:pack-propagation} illustrates this propagation.

Thus, by prioritizing \llm{} calls for functions that can unlock more subsequent
static inference, \tool{} can annotate all functions in the example with a
single \llm{} call.

\myparagraph{(3) Type-Guided Context Construction}
By leveraging the type analysis, we can construct context that is both effective and efficient.
The key idea is twofold: (1) precisely identifying relevant functions through type analysis and (2) providing type hints about incoming values.

Suppose that \tool{} selects the function |pack| at line~15 as the next \llm{} target.
For the call |k.kind(name)| in |pack| at line~18, type analysis determines that |k| has type
|String|, so \tool{} includes |String.kind| while omitting irrelevant functions
with the same lexical name, such as |Bytes.kind|.

Next, prompt size constraints often prevent a tool from including all related
caller and callee bodies.
To compensate, \tool{} provides type hints summarized from static analysis.
For example, we consider the caller context of |pack|, which includes |internal| and |external|.
However, to know that |paths| has type |list[str]|, 
a context construction method needs to include the function |get_paths| used by |internal|.
Instead of adding this extra function body to the prompt, \tool{} summarizes the types of values that flow into |paths|.
Using type analysis, \tool{} can determine that the call at line~7 passes a
|list[str]| value to |paths|, while the call at line~12 passes an unknown value.
Providing this summary conveys the useful evidence about |paths| without including additional function bodies in the prompt.
Figure~\ref{fig:context-ours} shows this context construction for function |pack|.



\section{Algorithm}
\label{sec:new_algorithm}

\myparagraph{Preliminaries}
We define a program $P \in \textsl{Program}$ as a set of function declarations.
A function declaration $\Func$ is defined as a tuple $(m, \ParamSeq)$, where $m$ is the function name and $\ParamSeq = (\Param_1, \ldots, \Param_n)$ is a sequence of parameters.
We write $\mathsf{Names}(P)$ for the set of function names declared in $P$.
We define a type domain $\Types$ as follows:
\[
  \Types = \textsl{BuiltInTypes} \cup \textsl{CustomTypes} \cup \{\Unknown{}\}
\]
where \textsl{BuiltInTypes} is the set of built-in types (e.g., |int|, |str|, |list|), \textsl{CustomTypes} is the set of user-defined types (e.g., |MyClass|), and \Unknown{} is a special element representing an unresolved type.
We write $t \in \Types$ for a single type and $T \subseteq \Types$ for a set of types.

\myparagraph{Problem}
Given an untyped program $P$, our goal is to compute an annotation map $\Ann$:
\[
  \Ann \in \textsl{AnnotMap}
  \qquad
  \Ann : \Slots{P} \rightarrow 2^{\Types}
\]
where \Slots{P} is the set of parameter type slots:
\[
  \Slots{P}
  =
  \{(\FuncName, i) \mid (\FuncName, \ParamSeq) \in P \text{ and } 1 \leq i \leq \lvert {\ParamSeq{}} \rvert \}.
\]
A slot $s=(\FuncName,i)$ denotes the $i$-th parameter of function $\FuncName$; return slots are handled analogously and omitted for simplicity.

\begin{algorithm}[t]
\caption{Overall algorithm of \textnormal{\tool{}}}
\label{fig:overall-algorithm}
\begin{algorithmic}[1]
\REQUIRE Program $P$, the number of queries $k$
\ENSURE Annotation map $\Ann$
\STATE $\Ann \gets [s \mapsto \{\Unknown{}\} \mid s\in\Slots{P}]$ \label{alg:init-ann}
\STATE $H \gets ([\,],[\,],[\,])$ \hfill (empty evidence maps)
\WHILE{$\mathsf{Unresolved}(P,\Ann) \neq \emptyset$} \label{alg:while}
\REPEAT \label{alg:repeat}
  \STATE $\Ann_{\mathit{old}}, H_{\mathit{old}} \gets \Ann, H$
  \STATE $H \gets \mathsf{TypeAnalysis}(P,\Ann)$ \label{alg:type-analysis} \hfill (Sec~\ref{sec:type_analysis})
  \STATE $\Ann \gets \mathsf{Update}(P,\Ann, H)$ \label{alg:update-ann}
\UNTIL{($\Ann, H) = (\Ann_{\mathit{old}}, H_{\mathit{old}})$} \label{alg:repeat-end}
  \IF{$\mathsf{Unresolved}(P,\Ann) = \emptyset$}
    \STATE \textbf{break} \label{alg:break}
  \ENDIF
  \STATE $M \gets \mathsf{TopFuncs}(P, \Ann, H, k)$ \label{alg:candidates} \hfill (Sec~\ref{sec:prioritizing})
  \FORALL{$m\in M$}
    \STATE $C \gets \mathsf{Ctx}(P, \Ann, H, m)$ \label{line:ctx} \hfill (Sec~\ref{sec:type_aware_context})
    \STATE $\Ann \gets \mathsf{QueryLLM}(C, \Ann, m)$ \label{alg:query-llm}
  \ENDFOR
\ENDWHILE \label{alg:end-while}
\RETURN $\Ann$ \label{alg:return}
\end{algorithmic}
\end{algorithm}

\myparagraph{Overall Algorithm}
Algorithm~\ref{fig:overall-algorithm} presents the overall workflow of \tool{}.
\tool{} first initializes every slot as \Unknown{} (line~\ref{alg:init-ann}).
As described in Section~\ref{sec:overview},
we first perform static type analysis to collect forward and backward evidence (line~\ref{alg:type-analysis}),
select high-leverage functions for \llm{} queries (line~\ref{alg:candidates}), and construct type-aware context for each selected function (line~\ref{line:ctx}).
We repeat this process until all slots are resolved (line~\ref{alg:while}--\ref{alg:end-while}).
We define $\mathsf{Unresolved}(P, \Ann)$ as the set of unresolved methods:
\[
\begin{aligned}
  &\mathsf{Unresolved}(P,\Ann)
  =\\
  &\{\, m \mid \exists i.\ (m,i) \in \Slots{P}
      \land \Ann((m,i))=\{\Unknown{}\} \,\}
\end{aligned}
\]
The following subsections detail each step.

\subsection{Forward and Backward Type Analysis}\label{sec:type_analysis}

To infer annotations for the slots in $\Ann$, \tool{} follows a two-step
process.
First, forward and backward analyses collect type candidates for each slot from
caller-side information and use-site evidence, respectively (line~\ref{alg:type-analysis}).
The result is an evidence tuple $H=(\FHint,\BStrong,\BWeak)$, where
$\FHint$ is produced by forward analysis and $\BStrong$ and $\BWeak$ are produced by backward analysis.
As illustrated in Listing~\ref{lst:combined-evidence},
backward evidence is split into strong evidence $\BStrong$, which can be used directly, 
and weak evidence $\BWeak$, which is informative but must be used conservatively.

Second, \tool{} combines these evidence maps to determine which annotations can be
added directly to $\Ann$ (line~\ref{alg:update-ann}).

\myparagraph{Forward Analysis}
Forward analysis collects caller-side type candidates in a map $\FHint$:
\[
  \FHint : \Slots{P} \rightarrow 2^{\Types}.
\]
For a slot $s=(m,i)$, $\FHint(s)$ summarizes the types of values that callers pass to the $i$-th
parameter of $m$.

\tool{} analyzes each function independently.
For a function $(\FuncName, \ParamSeq)$ with
$\ParamSeq=(\Param_1,\ldots,\Param_n)$, the analysis initializes a local type
environment $\Gamma_{\FuncName} : \Var \rightarrow 2^{\Types}$ from the annotation map:
\[
  \Gamma_{\FuncName}(\Param_i) = \Ann(s_i)
  \quad\text{where } s_i=(\FuncName,i) \text{ and } 1 \leq i \leq n.
\]
It then propagates types through standard typing rules, updating $\Gamma_{\FuncName}$ as it encounters assignments and expressions.

The key update happens at call sites.
When the analysis reaches a call site $\ell$, it uses the current local type environment to resolve the callee when possible.
If the callee is resolved to a function $m_\ell$ and there is an argument $a_i$ passed to the $i$-th parameter of $m_\ell$,
\tool{} records this caller-side evidence by updating:
\[
  \FHint(s_i^\ell) \leftarrow \FHint(s_i^\ell) \cup \Gamma_{\FuncName}(a_i)
  \quad\text{where } s_i^\ell=(m_\ell,i).
\]
This update is applied only when the analysis can identify the target function precisely enough.
The forward analysis continues until all functions have been analyzed, producing a complete map of caller-side type candidates.

\myparagraph{Backward Analysis}
Backward analysis collects use-site evidence in two maps:
\[
  \BStrong,\BWeak : \Slots{P} \rightarrow 2^{\Types}.
\]
\revise{A2}{
$\BStrong$ stores explicit evidence from direct type checks in the function body, such as \texttt{isinstance(x, t)}.
$\BWeak$ stores indirect evidence from callee parameter types, which describe accepted inputs rather than directly constraining the caller's argument type.
}

For each function $(\FuncName, \ParamSeq)$, \tool{} traverses paths from return statements
backward to the function entry.
It maintains two local evidence maps,
$\Gamma_{\FuncName}^{s}$ and $\Gamma_{\FuncName}^{w}$, which track strong and
weak evidence for local variables, respectively:
\[
  \Gamma_{\FuncName}^{s}, \Gamma_{\FuncName}^{w}
  : \Var \rightarrow 2^{\Types}.
\]
After the traversal finishes, \tool{} transfers the evidence accumulated for
each parameter $\Param_i$ to the corresponding global slot $s_i=(\FuncName,i)$
for $1 \leq i \leq \lvert \ParamSeq \rvert$:
\[
  \BStrong(s_i) \leftarrow \BStrong(s_i) \cup \Gamma_{\FuncName}^{s}(\Param_i)
  \quad
  \BWeak(s_i) \leftarrow \BWeak(s_i) \cup \Gamma_{\FuncName}^{w}(\Param_i).
\]



\example{}
Consider the following program:
\begin{lstlisting}[style=pythoncode, escapeinside={(*@}{@*)},belowskip=0.4em]
def g(z: int | str): ...
def f(x, y):
    z = x
    if isinstance(z, int):
        return z
    else:
        return g(y)
\end{lstlisting}

The guard |isinstance(z, int)| yields strong evidence for |z|, assigning $\mathsf{int}$ to $\Gamma_f^{s}(z)$.
Following the assignment from |z| back to the first parameter |x|, the analysis
records this evidence as $\BStrong(s_1^{f})=\{\mathsf{int}\}$, where
$s_1^{f}=(f,1)$.
In contrast, |g(y)| yields weak evidence for |y| because |g| expects an |int| or |str| value as its first argument.
The analysis therefore adds $\{\mathsf{int},\mathsf{str}\}$ to $\Gamma_f^{w}(y)$ and records it as
$\BWeak(s_2^{f})=\{\mathsf{int},\mathsf{str}\}$, where $s_2^{f}=(f,2)$.


\myparagraph{Type Annotation from Evidence}
After collecting evidence, \tool{} updates $\Ann$ with $\mathsf{Update}(P,\Ann,H)$ (line~\ref{alg:update-ann}):
\[
\begin{aligned}
  \mathsf{Update}(P, \Ann, H) = 
  \Ann[
    s \mapsto 
    \mathsf{Concrete}(s, \Ann, H)
    \mid s \in \Slots{P}
  ]
\end{aligned}
\]
Here, $\mathsf{Concrete}$ replaces $\Ann(s)$ only when the evidence yields a fully concrete type set; otherwise, it preserves $\Ann(s)$:
\[
\mathsf{Concrete}(s, \Ann, H) =
  \begin{cases}
    \mathbb{T}(s, H) & \text{if } \Unknown{} \notin \mathbb{T}(s, H)\\
    \Ann(s) & \text{otherwise}
  \end{cases}
\]
$\mathbb{T}(s, H)$ combines evidence to determine the final set of candidate types for slot $s$:
\[
\begin{aligned}
  & \mathbb{T}(s, (\FHint, \BStrong, \BWeak)) = \\
  & \begin{cases}
    \BStrong(s) \cup \{t\}
      & \text{if } \BWeak(s)=\{t\} \text{ and } \FHint(s) = \{ \Unknown{} \},\\
    \BStrong(s) \cup \FHint(s)
      & \text{otherwise}
  \end{cases}
\end{aligned}
\]
\revise{A2}{
For direct annotation, we use forward and strong backward evidence.
Weak backward evidence is used only when it yields a single candidate type and no forward evidence is available.
}

We repeat the type analysis and annotation update until no further changes occur (lines~\ref{alg:repeat}--\ref{alg:repeat-end}).

\subsection{Prioritizing LLM Calls}\label{sec:prioritizing}

\tool{} invokes the \llm{} for functions that still contain unresolved slots.
The order of these calls matters because an \llm{}-generated annotation can
restart static propagation and enable further static inference without additional \llm{} calls.

\tool{} selects the $k$ functions with the highest priority for \llm{} queries (line~\ref{alg:candidates}):
\[
\mathsf{TopFuncs}(P, \Ann, H, k) = \mathsf{Top}_k(\Pi(P, \Ann, H))
\]
where $\Pi(P, \Ann, H)$ associates each unresolved function with its priority scores:
\[
\begin{aligned}
  \Pi(P,\Ann,H)
  =
  \{ (m,
    \mathsf{Impact}(P, \Ann, m),
    \mathsf{Richness}(P, H,\Ann, m)
  ) & \\ 
  \mid m \in \mathsf{Unresolved}(P, \Ann) \} &
\end{aligned}
\]
$\mathsf{Impact}$ measures how many unresolved call-graph neighbors a function has, while $\mathsf{Richness}$ estimates the informativeness of its type hints.
Selecting the top-$k$ functions by these scores prioritizes calls likely to unlock static propagation and provide rich evidence.

\myparagraph{Impact Score}
The impact score of a function $m$ is defined as the number of unresolved functions that are directly connected to $m$ in the call graph.
\tool{} constructs a call graph $\mathcal{G}_{\Ann}$ using the current annotation map $\Ann$:
\[
  \mathcal{G}_{\Ann} \subseteq \textsl{Methods} \times \textsl{Methods},
\]
where $(m,n)\in \mathcal{G}_{\Ann}$ means that function $m$ calls function $n$.
The call graph can be incomplete if some types are still unresolved.
\tool{} intentionally omits such unresolved edges because speculative edges do not support static propagation, which is the objective of prioritizing \llm{} calls.

\tool{} calculates the impact score of a function $m$ as follows:
\[
  \begin{aligned}
  & \mathsf{Impact}(P, \Ann,m)
  = \\
  & \lvert
  \{\,n\in\mathsf{Unresolved}(P, \Ann)
    \mid (n,m)\in\mathcal{G}_{\Ann}
    \lor (m,n)\in\mathcal{G}_{\Ann}\,\}
  \rvert
  \end{aligned}
\]
A high score indicates annotating $m$ is likely to unlock forward propagation to callees or backward propagation to callers.

\myparagraph{Richness Score}
As a tie breaker, \tool{} uses a richness score that estimates the informativeness of type hints for the unresolved slots of $m$.
We define $\phi(P, \Ann,m)$ as the set of unresolved slots for function $m$:
\[
  \begin{aligned}
  \phi(P, \Ann,m)
  =
  \{\,(m,i) \mid {} &(m,i) \in \Slots{P} \\
    &{}\land \Ann((m,i)) = \{ \Unknown{} \}\,\}.
  \end{aligned}
\]
Then, the richness score is the average hint score over these unresolved slots:
\[
  \mathsf{Richness}(P, H, \Ann, m)
  =
  \frac{\sum_{s\in\phi(P, \Ann,m)}
    \mathsf{Hint}(H,s)}
  {\lvert\phi(P, \Ann,m)\rvert}.
\]
The function $\mathsf{Hint}$ assigns a higher score when both caller-side hints and
callee-side evidence contain concrete types, and a lower score when the evidence
is partial or unknown:
\[
\begin{aligned}
  & \mathsf{Hint}((\FHint, \BStrong, \BWeak), s) = \\
  & \left\{
  \begin{array}{r@{\quad}l@{\;}l}
  1.00 & \mathsf{IsConcrete}(\FHint(s)) & {}\land \mathsf{IsConcrete}(B(s))\\
  0.75 & \mathsf{IsConcrete}(\FHint(s)) & {}\land \mathsf{IsPartial}(B(s))\\
  0.75 & \mathsf{IsPartial}(\FHint(s)) & {}\land \mathsf{IsConcrete}(B(s))\\
  0.50 & \mathsf{IsPartial}(\FHint(s)) & {}\land \mathsf{IsPartial}(B(s))\\
  0.25 & \mathsf{IsPartial}(\FHint(s)) & {}\land B(s) =\{\Unknown{}\}\\
  0.00 & \text{otherwise} &
  \end{array}
  \right. \\ 
  & \text{where } B(s) = \BStrong(s) \cup \BWeak(s)
\end{aligned}
\]
The function $\mathsf{IsConcrete}(T)$ returns true if $T$ contains only concrete types, while $\mathsf{IsPartial}(T)$ returns true if $T$ contains at least one concrete type and at least one \Unknown{}.

The intuition is that clearer type evidence is more useful to the \llm{}.
We give slightly higher priority to slots with concrete forward hints, 
because weak backward evidence reflects callee-side expectations and
may lead the model to choose overly broad types.

\subsection{Type-Aware Context Construction}\label{sec:type_aware_context}
For each method selected by $\mathsf{TopFuncs}$, \tool{} builds a prompt from code context and type-hint context (line~\ref{line:ctx}):
\[
\begin{aligned}
  & \mathsf{Ctx}(P, \Ann, H, m)
  = \\
  & (
    \mathsf{CodeCtx}(\Ann, m),
    \{\mathsf{TypeHint}(s, H) \mid s \in \phi(P, \Ann, m)\}
  )
\end{aligned}
\]
where $\mathsf{CodeCtx}$ returns relevant code for $m$ and $\mathsf{TypeHint}$ returns evidence for slot $s$.

\myparagraph{Code Context}
Let $m$ be the target method. 
The code context includes caller and callee context resolved using the call graph $\mathcal{G}_{\Ann}$.
We define the caller and callee sets as follows:
\[
  \begin{aligned}
  \mathsf{Callers}(\Ann, m)
  &= \{n \mid (n,m)\in \mathcal{G}_{\Ann}\}\\
  \mathsf{Callees}(\Ann, m)
  &= \{n \mid (m,n)\in \mathcal{G}_{\Ann}\}
  \end{aligned}
\]
If the call graph is incomplete due to unresolved types, \tool{} falls back to lexical name matching to find additional callers and callees for the code context.

For callees, \tool{} includes the callee body only when the callee still has an unresolved signature; 
otherwise, its current signature is sufficient:
\[
  \mathsf{CalleeCtx}(n)
  =
  \begin{cases}
    \mathsf{Signature}(n) & \text{if } n \text{ is fully annotated}\\
    \mathsf{Body}(n) & \text{otherwise}\\
  \end{cases}
\]
where $\mathsf{Signature}(n)$ returns the signature of method $n$ and $\mathsf{Body}(n)$ returns the body of method $n$.

Thus, the code context for $m$ is the union of the target method, the caller bodies, and the callee contexts:
\[
  \begin{aligned}
  \mathsf{CodeCtx}(\Ann, m)
  =
  &\ \mathsf{Body}(m)\\
  &\cup \{ \mathsf{Body}(n) \mid n \in \mathsf{Callers}(\Ann, m) \} \\
  &\cup
  \{\mathsf{CalleeCtx}(n)
    \mid n\in\mathsf{Callees}(\Ann, m)\}
  \end{aligned}
\]

\myparagraph{Type-Hint Context}
The second part of the prompt summarizes the type evidence collected by static analysis. 
For each unresolved parameter slot $s \in \phi(P, \Ann, m)$ of the target method $m$,
\tool{} reports the candidate types suggested by caller-side propagation and callee-side evidence:
\[
  \mathsf{TypeHint}(s, (\FHint, \BStrong, \BWeak))
  =
  \bigl(s,\FHint(s),\BStrong(s),\BWeak(s)\bigr).
\]
Intuitively, $\FHint(s)$ describes types observed at call sites, $\BStrong(s)$
describes direct constraints from the body of $m$, and $\BWeak(s)$ describes
indirect constraints obtained from callees.

\myparagraph{Querying the LLM}
Finally, \tool{} queries the \llm{} with the constructed context to obtain type predictions for the unresolved slots of $m$ (line~\ref{alg:query-llm}).
\tool{} asks the \llm{} to predict a set of types for each unresolved slot, and it updates the annotation map $\Ann$ with the predictions.

\section{Evaluation}
\label{sec:evaluation}

\begin{table}[t]
\centering
\caption{Dataset statistics for the selected \typybench{} projects.}
\label{tab:dataset}
\begin{tabular}{@{}lrrrr@{}}
\toprule
\multicolumn{1}{c}{\textbf{Project}} &
\multicolumn{1}{c}{\textbf{Funcs}} &
\multicolumn{1}{c}{\textbf{Slots}} &
\multicolumn{1}{c}{\textbf{Files}} &
\multicolumn{1}{c}{\textbf{Lines}} \\
\midrule
black & 607 & 1,453 & 56 & 96,673 \\
fastapi &  274 & 1,717 & 53 & 11,076 \\
flake8 & 203 & 462 & 35 & 3,039 \\
flask & 326 & 786 & 23 & 4,008 \\
poetry & 865 & 1,825 & 144 & 14,012 \\
pre-commit & 328 & 909 & 61 & 5,193 \\
pc-hooks\footnotemark & 104 & 228 & 34 & 1,799 \\
private-gpt & 183 & 451 & 56 & 4,307 \\
rich & 866 & 2,372 & 81 & 19,960 \\
typer & 149 & 698 & 19 & 4,043 \\
urllib3 & 451 & 1,232 & 40 & 7,816 \\
vnpy & 410 & 851 & 31 & 4,216 \\
\midrule
Total & 4,766 & 12,984 & 633 & 176,142 \\
\bottomrule
\end{tabular}
\end{table}
\footnotetext{pre-commit-hooks}

\begin{table}[t]
\centering
\caption{\llm{} models used in our evaluation. Prices are shown as dollars per million input/output tokens.}
\label{tab:llm-models}
\setlength{\tabcolsep}{3pt}
\begin{tabular}{@{}lllll@{}}
\toprule
\textbf{Model} & \textbf{Params} & \textbf{Source} & \textbf{Serving} & \textbf{Price} \\
\midrule
Gemma 4~\cite{gemma4} & E4B-IT & Google & Local & -- \\
gpt-oss~\cite{gptoss} & 120B & OpenAI & Cerebras API & \$0.35/\$0.75 \\
GLM-4.7~\cite{glm47} & 358B & Z.ai & Cerebras API & \$2.25/\$2.75 \\
GPT-5.5 & -- & OpenAI & OpenAI API & \$5.00/\$30.00 \\
\bottomrule
\end{tabular}
\vspace{-1em}
\end{table}

\begin{table*}[t]
\centering
\caption{The accuracy and \llm{} usage on \typybench{}.
\textbf{Calls}: the number of \llm{} queries.
\textbf{Accuracy}: the percentage of correctly inferred type annotations.
\textbf{Time}: the total time taken for type inference.
\textbf{Input}: the total number of input tokens.
\textbf{Output}: the total number of output tokens.
\textbf{Price}: the estimated cost of \llm{} usage.
}
\label{tab:accuracy-results}

\begin{tabular}{@{}llrrrrrr@{}}
\toprule
\multicolumn{1}{c}{\multirow{2.5}{*}{\textbf{System}}} &
\multicolumn{1}{c}{\multirow{2.5}{*}{\textbf{Model}}} &
\multicolumn{1}{c}{\multirow{2.5}{*}{\textbf{Accuracy}}} &
\multicolumn{1}{c}{\multirow{2.5}{*}{\makecell{\textbf{Time}\\\textbf{(hh:mm)}}}} &
\multicolumn{4}{c}{\textbf{\llm{} Usage}} \\
\cmidrule(l){5-8}
& & & & \multicolumn{1}{c}{\textbf{Calls}} & \multicolumn{1}{c}{\textbf{Input (M)}} & \multicolumn{1}{c}{\textbf{Output (M)}} & \multicolumn{1}{c}{\textbf{Price (\$)}} \\
\midrule
\tiger{} & CodeT5 (Fine-tuned) & 49.83\% & 17:22 & -- & -- & -- & -- \\
\typetfive{} & CodeT5 (Fine-tuned) & 69.30\% & 01:53 & -- & -- & -- & -- \\
\typegen{} & gpt-oss & 46.51\% & 08:00 & 13,702 & 21.48 & 4.37 & \$10.79 \\
\midrule
\aider{} & Gemma 4 & 72.03\% & 26:12 & 5,507 & 126.83 & 1.09 & -- \\
\ours{} & Gemma 4 & ($\uparrow$6.6\%) \textbf{76.76\%} & ($\downarrow$32.6\%) \textbf{17:40} & ($\downarrow$40.1\%) \textbf{3,300} & ($\downarrow$79.0\%) \textbf{26.63} & ($\downarrow$69.9\%) \textbf{0.33} & -- \\
\midrule
\aider{} & gpt-oss & 82.41\% & 15:12 & 4,733 & 95.33 & 2.52 & \$35.25 \\
\ours{} & gpt-oss & ($\uparrow$2.0\%) \textbf{84.06\%} & ($\downarrow$6.0\%) \textbf{14:17} & ($\downarrow$33.1\%) \textbf{3,167} & ($\downarrow$76.1\%) \textbf{22.78} & ($\downarrow$14.2\%) \textbf{2.16} & ($\downarrow$72.8\%) \textbf{\$9.59} \\
\midrule
\aider{} & GLM-4.7 & 84.13\% & 16:15 & 4,752 & 119.16 & 3.77 & \$285.08 \\
\ours{} & GLM-4.7 & ($\uparrow$2.0\%) \textbf{85.82\%} & ($\downarrow$4.4\%) \textbf{15:32} & ($\downarrow$33.8\%) \textbf{3,146} & ($\downarrow$80.9\%) \textbf{22.77} & ($\downarrow$41.2\%) \textbf{2.22} & ($\downarrow$78.5\%) \textbf{\$61.19} \\
\bottomrule
\end{tabular}
\end{table*}

\begin{table*}[t]
\centering
\caption{Per-project comparison between \aider{} and \ours{} on \typybench{}.
\textbf{Accuracy}: the percentage of correctly inferred type annotations.
\textbf{Calls}: the number of \llm{} queries.
\textbf{Tokens}: the total number of input and output tokens.
}
\label{tab:per-project-results}
\setlength{\tabcolsep}{2pt}
\begin{tabular}{@{}lrrrrrrrrrrrrrrrrrr@{}}
\toprule
\multicolumn{1}{c}{\multirow{3.5}{*}[-0.35ex]{\textbf{Project}}} &
\multicolumn{6}{c}{\textbf{Gemma 4}} &
\multicolumn{6}{c}{\textbf{gpt-oss}} &
\multicolumn{6}{c}{\textbf{GLM-4.7}} \\
\cmidrule(lr){2-7} \cmidrule(lr){8-13} \cmidrule(l){14-19}
& \multicolumn{2}{c}{\textbf{Accuracy}} &
\multicolumn{2}{c}{\textbf{Calls}} &
\multicolumn{2}{c}{\textbf{Tokens (M)}} &
\multicolumn{2}{c}{\textbf{Accuracy}} &
\multicolumn{2}{c}{\textbf{Calls}} &
\multicolumn{2}{c}{\textbf{Tokens (M)}} &
\multicolumn{2}{c}{\textbf{Accuracy}} &
\multicolumn{2}{c}{\textbf{Calls}} &
\multicolumn{2}{c}{\textbf{Tokens (M)}} \\
\cmidrule(lr){2-3} \cmidrule(lr){4-5} \cmidrule(lr){6-7}
\cmidrule(lr){8-9} \cmidrule(lr){10-11} \cmidrule(lr){12-13}
\cmidrule(lr){14-15} \cmidrule(lr){16-17} \cmidrule(l){18-19}
& \multicolumn{1}{c}{\aider{}} & \multicolumn{1}{c}{\ours{}}
& \multicolumn{1}{c}{\aider{}} & \multicolumn{1}{c}{\ours{}}
& \multicolumn{1}{c}{\aider{}} & \multicolumn{1}{c}{\ours{}}
& \multicolumn{1}{c}{\aider{}} & \multicolumn{1}{c}{\ours{}}
& \multicolumn{1}{c}{\aider{}} & \multicolumn{1}{c}{\ours{}}
& \multicolumn{1}{c}{\aider{}} & \multicolumn{1}{c}{\ours{}}
& \multicolumn{1}{c}{\aider{}} & \multicolumn{1}{c}{\ours{}}
& \multicolumn{1}{c}{\aider{}} & \multicolumn{1}{c}{\ours{}}
& \multicolumn{1}{c}{\aider{}} & \multicolumn{1}{c}{\ours{}} \\
\midrule
black & 67.79\% & \textbf{71.16\%} & 649 & \textbf{418} & 18.98 & \textbf{3.15} & 78.05\% & \textbf{79.49\%} & 555 & \textbf{398} & 14.37 & \textbf{2.81} & \textbf{79.97\%} & 79.90\% & 558 & \textbf{396} & 15.01 & \textbf{2.83} \\
fastapi & 53.58\% & \textbf{54.63\%} & 336 & \textbf{238} & 8.57 & \textbf{2.03} & 65.06\% & \textbf{74.61\%} & 269 & \textbf{217} & 6.06 & \textbf{1.80} & 76.70\% & \textbf{79.21\%} & 281 & \textbf{212} & 6.43 & \textbf{1.76} \\
flake8 & 81.39\% & \textbf{85.28\%} & 232 & \textbf{126} & 2.84 & \textbf{0.56} & 84.42\% & \textbf{87.01\%} & 205 & \textbf{122} & 2.30 & \textbf{0.56} & 83.77\% & \textbf{87.66\%} & 204 & \textbf{122} & 2.36 & \textbf{0.55} \\
flask & \textbf{66.67\%} & 63.99\% & 393 & \textbf{286} & 9.26 & \textbf{1.92} & \textbf{78.24\%} & 76.21\% & 330 & \textbf{262} & 7.10 & \textbf{1.74} & 79.77\% & \textbf{80.41\%} & 328 & \textbf{259} & 7.14 & \textbf{1.75} \\
poetry & 81.26\% & \textbf{85.15\%} & 995 & \textbf{579} & 23.54 & \textbf{4.54} & \textbf{90.19\%} & 89.81\% & 871 & \textbf{557} & 18.76 & \textbf{4.27} & 90.14\% & \textbf{90.58\%} & 872 & \textbf{560} & 25.49 & \textbf{4.31} \\
pre-commit & 80.75\% & \textbf{86.91\%} & 371 & \textbf{220} & 6.54 & \textbf{0.83} & 90.32\% & \textbf{90.98\%} & 332 & \textbf{205} & 5.16 & \textbf{0.81} & 84.82\% & \textbf{90.87\%} & 331 & \textbf{206} & 5.25 & \textbf{0.80} \\
pc-hooks & 90.35\% & \textbf{91.67\%} & 105 & \textbf{45} & 0.68 & \textbf{0.12} & 93.86\% & \textbf{94.74\%} & 105 & \textbf{44} & 0.63 & \textbf{0.13} & 92.98\% & \textbf{94.74\%} & 101 & \textbf{44} & 0.63 & \textbf{0.13} \\
private-gpt & 80.27\% & \textbf{87.80\%} & 219 & \textbf{151} & 3.46 & \textbf{0.93} & 89.14\% & \textbf{89.58\%} & 187 & \textbf{150} & 2.65 & \textbf{0.91} & 89.58\% & \textbf{91.13\%} & 188 & \textbf{150} & 2.69 & \textbf{0.89} \\
rich & 74.33\% & \textbf{84.06\%} & 1,035 & \textbf{577} & 28.74 & \textbf{7.70} & 87.69\% & \textbf{88.07\%} & 871 & \textbf{566} & 21.52 & \textbf{7.13} & 89.04\% & \textbf{90.56\%} & 885 & \textbf{560} & 37.92 & \textbf{7.23} \\
typer & 59.03\% & \textbf{63.75\%} & 181 & \textbf{109} & 2.60 & \textbf{0.74} & 78.37\% & \textbf{78.94\%} & 156 & \textbf{99} & 1.96 & \textbf{0.64} & 78.94\% & \textbf{80.09\%} & 149 & \textbf{100} & 1.90 & \textbf{0.65} \\
urllib3 & 69.97\% & \textbf{77.52\%} & 510 & \textbf{270} & 13.79 & \textbf{2.34} & 79.30\% & \textbf{81.25\%} & 439 & \textbf{273} & 10.63 & \textbf{2.20} & 79.30\% & \textbf{82.55\%} & 441 & \textbf{263} & 11.40 & \textbf{2.14} \\
vnpy & 85.31\% & \textbf{88.60\%} & 481 & \textbf{281} & 8.92 & \textbf{2.10} & \textbf{88.95\%} & 88.13\% & 413 & \textbf{274} & 6.70 & \textbf{1.95} & \textbf{89.19\%} & 88.72\% & 414 & \textbf{274} & 6.73 & \textbf{1.94} \\
\midrule
Overall & 72.03\% & \textbf{76.76\%} & 5,507 & \textbf{3,300} & 127.92 & \textbf{26.95} & 82.41\% & \textbf{84.06\%} & 4,733 & \textbf{3,167} & 97.85 & \textbf{24.94} & 84.13\% & \textbf{85.82\%} & 4,752 & \textbf{3,146} & 122.93 & \textbf{24.98} \\
\bottomrule
\end{tabular}
\vspace{-1em}
\end{table*}

We evaluate \tool{} through the following research questions:
\begin{itemize}
  \item \textbf{RQ1. Effectiveness and Efficiency:} How does \tool{} compare to type inference baselines in terms of annotation accuracy and \llm{} usage?
  \item \textbf{RQ2. Generalization to a Frontier LLM:} Does \tool{} remain effective when paired with a frontier \llm{}?
  \item \textbf{RQ3. Ablation Study:} How do different components of \tool{} contribute to its performance?
  \item \textbf{RQ4. Case Study:} What are representative examples where \tool{} succeeds or fails?
\end{itemize}

\subsection{Experimental Setup}
\label{sec:setup}

\tool{} is implemented in about 15,000 lines of Python 3.13.12 code on top of \aider{} (v.0.86.3) \cite{aider}.
\tool{} uses \jedi{} (v.0.20.0) \cite{jedi} for call resolution and \pyrefly{} (v.1.0.0) \cite{pyrefly} for return type inference.
We run all experiments on a Linux machine (Debian 13) with an Intel Xeon Silver 4214 CPU, an NVIDIA RTX A6000 GPU, and 128 GB of memory.

\myparagraph{Dataset}
\revise{A3,B2}{
We conduct a repository-level evaluation on 12 high-quality projects from \typybench{} \cite{typybench}, removing all type annotations and inferring types for all functions in each project.
}
We select projects that satisfy the following criteria:
at least 3,000 GitHub stars, at least 95\% function-level annotation coverage, and between 100 and 1,000 functions to keep the number of \llm{} API calls and evaluation cost manageable.
Table~\ref{tab:dataset} summarizes the selected projects, which include diverse domains such as web frameworks, code formatters, and data processing libraries.
Next, we remove all type annotations from these projects to create an untyped codebase for evaluation.
We compute accuracy over normalized type annotations,
since the same Python type can be written in multiple syntactic forms, such as |Union[int, None]| and |Optional[int]|.


\myparagraph{LLM Models}
For RQ1, we evaluate \tool{} with Gemma 4, gpt-oss, and GLM-4.7; the RQ3 ablations use gpt-oss.
Gemma 4 runs on a local machine, while gpt-oss and GLM-4.7 are accessed through the Cerebras API\footnote{\url{https://www.cerebras.ai}}.
For RQ2, we use GPT-5.5 through the OpenAI API as a frontier \llm{}.
Table~\ref{tab:llm-models} summarizes the model size, serving environment, and pricing for each model.

We use temperature 0 for all models except GLM-4.7.
For GLM-4.7, we set temperature 1 because temperature 0 frequently causes non-terminating responses in our experiments.
When an \llm{} response is not parseable, we query the model again, allowing at most two retries.
If the response still cannot be parsed, we skip the function.

\myparagraph{Baselines}
We compare \tool{} against the following baselines:
\begin{itemize}
    \item \textbf{\tiger{}} \cite{tiger}: a neural type inference system that uses a graph neural network to predict types.
    \item \textbf{\typetfive{}} \cite{typet5}: a sequence-to-sequence model that generates type annotations.
    \item \textbf{\typegen{}} \cite{typegen}: an \llm{}-based system that generates type annotations.
    \item \textbf{\aider{}} \cite{aider}: a general-purpose \llm{} coding assistant that can be prompted to add type annotations.
\end{itemize}
For the end-to-end evaluation, we exclude \hityper{} \cite{hityper} and \nester{} \cite{nester} because we could not obtain runnable prediction pipelines from their released artifacts.
In our attempts, the Type4Py-based Docker server \cite{similar-type4py} used by \hityper{} failed during installation, and the artifact for \nester{} appeared to require external files that were not included.

Thus, we select \tiger{}, \typetfive{}, and \typegen{} as reproducible state-of-the-art specialized baselines in Python, 
covering recent approaches that predict individual type slot annotations, infer complete function signatures, and use \llm{}-based generation.
In addition, we use \aider{} as the general-purpose coding assistant baseline due to its open-source, model-agnostic, and scriptable design.
This allows us to build \tool{} on the same infrastructure and isolate the effect of our techniques under controlled, multi-model experiments.

For each baseline, we follow the prediction granularity used in its original evaluation.
\tiger{} and \typegen{} predict one type at a time for each argument or return slot.
In contrast, \typetfive{} and \aider{} infer all missing annotations in a target function in a single prediction step.

For an ablation study, we also compare \tool{} against two static type inference tools: \hityper{} \cite{hityper} and \pyrefly{} \cite{pyrefly}.
Although \hityper{} does not provide access to its language model,
its static analysis component can be run independently to produce type annotations.
In addition, we include \pyrefly{} as a representative industrial type inference tool.

\subsection{RQ1. Effectiveness and Efficiency}

Table~\ref{tab:accuracy-results} compares the accuracy and \llm{} usage of \tool{} against the baselines on the benchmark.
\aider{} outperforms the specialized type inference tools, making it the strongest baseline in our evaluation.
\tool{} further improves over \aider{}, achieving higher accuracy for all evaluated model families.
The gain is most pronounced for the smaller local model, where accuracy improves from 72.03\% to 76.76\%.

The results also show that the accuracy gains do not come from using more \llm{} calls.
Instead, \tool{} substantially reduces model usage by resolving types statically and querying the \llm{} for high-leverage targets.
With gpt-oss, the number of calls decreases by 33.1\% and the estimated price drops from \$35.25 to \$9.59, a 72.8\% reduction.
With GLM-4.7, the estimated price decreases from \$285.08 to \$61.19, a 78.5\% reduction.
The runtime benefit is largest for Gemma 4, where time decreases from 26:12 to 17:40.
This suggests that static analysis is especially useful when model serving is relatively slow, such as on a local machine.

\revise{B3}{
Table~\ref{tab:per-project-results} presents a per-project comparison.
To assess whether \tool{} outperforms \aider{}, we apply Wilcoxon signed-rank tests \cite{wilcoxon} to the paired project-level results.
\tool{} reduces \llm{} calls and token usage in all 36 project--model pairs, with statistically significant reductions for each model ($p<0.001$ for both metrics).
For accuracy, the p-values are $0.001$, $0.055$, and $0.002$ for Gemma 4, gpt-oss, and GLM-4.7, respectively, indicating significant improvements for Gemma 4 and GLM-4.7.
Overall, \tool{} achieves consistent cost savings and generally improved accuracy.
}

\subsection{RQ2. Generalization to a Frontier LLM}
\label{sec:frontier-evaluation}

\begin{table}[t]
\centering
\caption{Per-project comparison between \aider{} and \ours{} with GPT-5.5 on the 10\% sampled-file setting.
\textbf{Sample Slots}: the number of evaluated annotation slots in the sampled files.
\textbf{Accuracy}: the percentage of correctly inferred type annotations.
\textbf{Calls}: the number of \llm{} queries.
\textbf{Price}: estimated with \$5/M input tokens and \$30/M output tokens.
}
\label{tab:gpt55-sample-results}
\setlength{\tabcolsep}{3pt}
\begin{tabular}{@{}lrrrrrrr@{}}
\toprule
\multicolumn{1}{c}{\multirow{2.5}{*}{\textbf{Project}}} &
\multicolumn{1}{c}{\multirow{2.5}{*}{\makecell{\textbf{Sample}\\\textbf{Slots}}}} &
\multicolumn{2}{c}{\textbf{Accuracy}} &
\multicolumn{2}{c}{\textbf{Calls}} &
\multicolumn{2}{c}{\textbf{Price (\$)}} \\
\cmidrule(lr){3-4} \cmidrule(lr){5-6} \cmidrule(l){7-8}
& & \aider{} & \ours{} & \aider{} & \ours{} & \aider{} & \ours{} \\
\midrule
black & 325 & 89.23\% & \textbf{90.77\%} & 119 & \textbf{93} & 21.43 & \textbf{5.38} \\
fastapi & 52 & \textbf{94.23\%} & \textbf{94.23\%} & 12 & \textbf{11} & 1.46 & \textbf{0.60} \\
flake8 & 115 & 92.17\% & \textbf{93.04\%} & 48 & \textbf{29} & 3.24 & \textbf{0.85} \\
flask & 160 & 88.12\% & \textbf{88.75\%} & 63 & \textbf{44} & 7.91 & \textbf{1.72} \\
poetry & 160 & \textbf{98.12\%} & 96.25\% & 90 & \textbf{57} & 10.25 & \textbf{1.96} \\
pre-commit & 130 & \textbf{96.15\%} & \textbf{96.15\%} & 52 & \textbf{34} & 4.53 & \textbf{0.70} \\
pc-hooks & 30 & 90.00\% & \textbf{96.67\%} & 13 & \textbf{5} & 0.44 & \textbf{0.08} \\
private-gpt & 38 & 97.37\% & \textbf{100.00\%} & 15 & \textbf{13} & 1.08 & \textbf{0.29} \\
rich & 82 & \textbf{100.00\%} & 97.56\% & 34 & \textbf{23} & 3.54 & \textbf{1.48} \\
typer & 133 & \textbf{90.23\%} & 87.97\% & 29 & \textbf{25} & 2.64 & \textbf{1.34} \\
urllib3 & 138 & \textbf{94.93\%} & 92.75\% & 61 & \textbf{44} & 7.49 & \textbf{1.79} \\
vnpy & 243 & 92.18\% & \textbf{95.06\%} & 104 & \textbf{70} & 9.41 & \textbf{2.31} \\
\midrule
Overall & 1,606 & 92.71\% & \textbf{93.09\%} & 640 & \textbf{448} & 73.42 & \textbf{18.49} \\
\bottomrule
\end{tabular}
\vspace{-1em}
\end{table}

To evaluate whether \tool{} remains effective with a stronger frontier \llm{}, we additionally compare \tool{} and \aider{} using GPT-5.5.
Due to the high cost of frontier-model evaluation, we sample 10\% of files from each project, remove type annotations only from the sampled files, and evaluate both systems on these files.

Table~\ref{tab:gpt55-sample-results} shows that \tool{} continues to provide benefits even with a much stronger model.
In the sampled-file setting, \tool{} improves accuracy from 92.71\% to 93.09\% while reducing \llm{} calls from 640 to 448.
The cost reduction is more substantial: the estimated price decreases from \$73.42 to \$18.49, a 74.8\% reduction, on the 10\% sampled files.
These results suggest that static analysis remains useful even when the underlying \llm{} is highly capable, primarily by avoiding unnecessary calls and reducing the amount of context sent to the model.

\subsection{RQ3. Ablation Study}
\begin{table}[t]
\centering
\caption{Ablation results with gpt-oss on the 12 projects.
\textbf{Accuracy}: the percentage of correctly inferred type annotations.
\textbf{Tokens}: the total number of input and output tokens.
\textbf{Calls}: the number of \llm{} queries.
\textbf{Price}: the estimated API cost at \$0.35/M input tokens and \$0.75/M output tokens.
}
\label{tab:ablation}
\small
\setlength{\tabcolsep}{3pt}
\begin{tabular}{@{}lrrrr@{}}
\toprule
\textbf{Configuration} & \textbf{Accuracy} & \textbf{Calls} & \textbf{Tokens (M)} & \textbf{Price (\$)} \\
\midrule
\textsc{ForwardOnly} & 81.16\% & 3,666 & 29.00 & 11.18 \\
\textsc{NoStaticFill} & 83.06\% & 5,069 & 36.83 & 14.05 \\
\textsc{NoFeedback} & 83.00\% & 3,814 & 31.50 & 12.07 \\
\midrule
\textsc{CalleeFirst} & 82.45\% & 3,373 & 26.79 & 10.31 \\
\textsc{NoCompression} & 83.31\% & 3,554 & 74.10 & 26.89 \\
\midrule
\ours{} & \textbf{84.06\%} & \textbf{3,167} & \textbf{24.94} & \textbf{9.59} \\
\bottomrule
\end{tabular}
\end{table}

\revise{B3}{
We evaluate five ablated configurations against the full \tool{} using gpt-oss on the 12 projects.

\myparagraph{Contribution of Static Analysis}
We examine the roles of backward inference, direct static annotation, and feedback propagation through three configurations:
\begin{itemize}
  \item \textbf{\textsc{ForwardOnly}} uses \pyrefly{}~\cite{pyrefly} for forward type analysis without backward inference.
  \item \textbf{\textsc{NoStaticFill}} disables direct annotation by static analysis while retaining static hints and prioritization.
  \item \textbf{\textsc{NoFeedback}} performs the initial static analysis and filling, then queries the remaining functions without propagating subsequent \llm{} predictions through static analysis.
\end{itemize}
Compared with the full \tool{} configuration, \textsc{ForwardOnly} has the largest accuracy drop, from 84.06\% to 81.16\%.
\textsc{NoStaticFill} increases calls by 60.1\%, highlighting the benefit of direct static annotation.
\textsc{NoFeedback} increases calls by 20.4\%, showing that propagating \llm{} predictions further reduces the query workload.

\myparagraph{Contribution of LLM Query Design}
We examine query prioritization and context compression through two configurations:
\begin{itemize}
  \item \textbf{\textsc{CalleeFirst}} replaces our query prioritization with a callee-first order.
  \item \textbf{\textsc{NoCompression}} uses the default context of \aider{} without omitting irrelevant code or replacing supporting code with static hints.
\end{itemize}
\textsc{CalleeFirst} lowers accuracy from 84.06\% to 82.45\% and requires 6.5\% more \llm{} calls than \tool{}.
This result suggests that our query prioritization benefits both prediction accuracy and query efficiency.
\textsc{NoCompression} uses nearly three times as many tokens as the full \tool{} configuration, highlighting how context compression using type information reduces query costs.
}

\myparagraph{Static Annotation Quality}
To demonstrate the quality of our static analysis component,
we evaluate its effectiveness before invoking the \llm{}.
Specifically, we run the static analysis once on the initial untyped codebase
and measure both recall and precision.

\begin{table}[t]
\centering
\caption{Recall and precision of static type inference tools.
\textbf{Rec.}: the percentage of all annotation slots correctly resolved by static analysis.
\textbf{Prec.}: the percentage of correctly inferred type annotations among all inferred annotations.
}
\label{tab:static-results}
\setlength{\tabcolsep}{5pt}
\begin{tabular}{@{}lrrrrrr@{}}
\toprule
\multicolumn{1}{c}{\multirow{2.5}{*}{\textbf{Project}}} &
\multicolumn{2}{c}{\textbf{\hityper{}}} &
\multicolumn{2}{c}{\textbf{\pyrefly{}}} &
\multicolumn{2}{c}{\textbf{\tool{}}} \\
\cmidrule(lr){2-3} \cmidrule(lr){4-5} \cmidrule(l){6-7}
& \multicolumn{1}{c}{\textbf{Rec.}} & \multicolumn{1}{c}{\textbf{Prec.}} &
\multicolumn{1}{c}{\textbf{Rec.}} & \multicolumn{1}{c}{\textbf{Prec.}} &
\multicolumn{1}{c}{\textbf{Rec.}} & \multicolumn{1}{c}{\textbf{Prec.}} \\
\midrule
black & 8.74\% & 60.19\% & 25.12\% & 92.64\% & \textbf{30.35\%} & \textbf{96.29\%} \\
fastapi & \textbf{15.14\%} & 23.70\% & 5.77\% & 83.19\% & 8.79\% & \textbf{87.79\%} \\
flake8 & 17.32\% & 57.14\% & 29.22\% & 84.91\% & \textbf{39.83\%} & \textbf{96.84\%} \\
flask & 13.23\% & 44.83\% & \textbf{22.90\%} & 83.33\% & 22.52\% & \textbf{90.31\%} \\
poetry & 20.55\% & 60.29\% & 33.42\% & 93.99\% & \textbf{34.41\%} & \textbf{96.47\%} \\
pre-commit & 16.50\% & 49.50\% & 23.43\% & 92.61\% & \textbf{34.43\%} & \textbf{99.05\%} \\
pc-hooks & 32.02\% & 57.94\% & 36.84\% & 92.31\% & \textbf{64.47\%} & \textbf{99.32\%} \\
private-gpt & 23.50\% & 55.50\% & \textbf{30.38\%} & 92.57\% & 26.39\% & \textbf{95.97\%} \\
rich & 24.45\% & 50.88\% & \textbf{29.09\%} & 86.57\% & 28.08\% & \textbf{94.74\%} \\
typer & 11.03\% & 43.75\% & 12.75\% & 81.65\% & \textbf{15.62\%} & \textbf{87.90\%} \\
urllib3 & 24.27\% & 44.49\% & \textbf{26.87\%} & 85.75\% & 25.97\% & \textbf{94.96\%} \\
vnpy & 37.49\% & 76.87\% & \textbf{39.13\%} & 94.33\% & 33.14\% & \textbf{98.26\%} \\
\midrule
Overall & 19.64\% & 47.89\% & 25.15\% & 89.45\% & \textbf{27.24\%} & \textbf{95.44\%} \\
\bottomrule
\end{tabular}
\vspace{-1em}
\end{table}

Table~\ref{tab:static-results} compares the recall and precision of the static type inference in \tool{} with those of the \hityper{} and \pyrefly{} baselines.
\tool{} achieves the highest recall and precision, correctly resolving 27.24\% of all slots with 95.44\% precision.
This shows that \tool{} can infer a substantial number of correct type annotations through static analysis.
At the same time, the limited recall indicates that static analysis alone cannot resolve most annotation slots, motivating the need for \llm{} queries.

\revise{A1}{
\myparagraph{Type Hint Quality}
\label{sec:type-hint-quality}
We investigate the usefulness of static type hints as context for the \llm{}.
We measure the caller and callee code that the \llm{} would need to inspect to infer types of parameters without static hints.
This code averages about 1,830 tokens per parameter, compared with 26 tokens for a static hint.
Thus, static type hints reduce the token cost of supplying type information.

Among the hinted slots across the three models in RQ1, 47.6\% receive hints that exactly match the reference type.
On these slots, \tool{} achieves 9.5\% higher exact-match accuracy than \aider{}.
Incorrect hints account for 44.9\% of hinted slots, where \tool{}'s accuracy is 3.0\% lower than \aider{}'s on average.
Overall, \tool{} achieves an average relative accuracy gain of 4.1\% over \aider{} on hinted slots, suggesting that static hints generally provide useful context for type prediction.
}

\subsection{RQ4. Case Study}

\myparagraph{Strengths of \tool{}}
To understand the accuracy gains of \tool{}, we inspect the GLM-4.7 cases where \tool{} correctly predicts 678 slots missed by \aider{}.
Among 100 sampled slots, 54 involve complex generic or class types that often require evidence beyond the target function.
We find two recurring failure modes: the evidence is either outside the retrieved context or present but not used correctly by the model.

In Figure~\ref{fig:case-generic}, 
suppose the target function is |get_definitions| at line 7, which takes a parameter |fields|.
\tool{} propagates the return type of |get_fields| at line 1 through |get_openapi| at line 3 to infer |fields: list[ModelField]| in |get_definitions|.
By contrast, \aider{} does not retrieve |get_fields| because the evidence is one step beyond the direct caller, and predicts the broader |Sequence[ModelField]| from the loop over |fields| at line 8.

In Figure~\ref{fig:case-class}, 
let the target function be |publish| at line 7, which takes a parameter |cert|.
\tool{} observes that |handle| at line 1 passes either a |Path| object or |None| to |Publisher.publish|, so this caller-side evidence is propagated to the parameter |cert|.
Even though \aider{} has access to the caller |handle| in its context,
its reasoning trace describes |cert| as a ``certificate file path'' and concludes that it ``could be Union[str, None]''.
This suggests that the \llm{} is biased toward the common convention that certificate paths are represented as strings, even when the caller provides more specific evidence.
In contrast, static analysis preserves that concrete evidence across call boundaries.

\begin{figure}[t]
\centering
\begin{subfigure}{\linewidth}
\begin{lstlisting}[style=pythoncode]
def get_fields() -> list[ModelField]: ...

def get_openapi():
  all_fields = get_fields()
  get_definitions(all_fields)

def get_definitions(fields):
  for field in fields: ...
\end{lstlisting}
\caption{A FastAPI example. The target function is \texttt{get\_definitions}. While \tool{} infers \texttt{list[ModelField]}, \aider{} predicts \texttt{Sequence[ModelField]} for parameter \texttt{fields}.}
\vspace{0.5em}
\label{fig:case-generic}
\end{subfigure}
\vspace{0.5em}
\begin{subfigure}{\linewidth}
\begin{lstlisting}[style=pythoncode]
def handle():
  publisher = Publisher()
  cert = Path(...) if ... else None
  publisher.publish(cert)

class Publisher:
  def publish(self, cert=None):
    certificates = external_func()
    resolved_cert = cert or certificates.cert
\end{lstlisting}
\caption{A poetry example. The target function is \texttt{publish}. While \tool{} infers \texttt{Union[Path, None]}, \aider{} predicts \texttt{Union[str, None]} for parameter \texttt{cert}.}
\label{fig:case-class}
\end{subfigure}
\vspace{-1.5em}
\caption{Examples where \tool{} succeeds but \aider{} fails.}
\vspace{-1.0em}
\label{fig:case-strength}
\end{figure}

\myparagraph{Weaknesses of \tool{}}
Because \tool{} relies on static analysis to guide \llm{} annotation, incorrect static evidence can limit its effectiveness.
Inspecting the static-analysis errors in Table~\ref{tab:static-results}, we find that 121 of 169 are over-narrowing errors: 
71 infer a return type narrower than the developer-written annotation, and 50 infer a parameter type narrower than intended from observed call arguments.

Figure~\ref{fig:case-return-narrowing} illustrates over-narrowing in return types.
The function body strongly suggests |list[PackageNode]|: the implementation constructs a list, appends |PackageNode| objects, and returns it directly.
However, the developer-written annotation is |Sequence[PackageNode]|, intentionally exposing a more general interface.
This suggests that static evidence can be too concrete when the desired annotation intentionally abstracts over the implementation.

Figure~\ref{fig:case-param-narrowing} shows over-narrowing for parameters.
Within the observed project code, |border_style| is passed only as a string, so the static analysis infers |str|.
The ground truth, however, is |StyleType| because |Table| is part of a public API that accepts both |Style| objects and strings.
This limitation is inherent to evidence-driven static inference: when all observed uses are narrow, the analysis can miss broader types that reflect external API intent rather than internal usage.


\begin{figure}[t]
\centering
\begin{subfigure}{\linewidth}
\begin{lstlisting}[style=pythoncode]
def reachable(self):
  children = []
  for d in self.all_requires:
    children.append(PackageNode(d))
  return children
\end{lstlisting}
\caption{Return type over-narrowing. The static analysis infers \texttt{list[PackageNode]}, while the developer-written annotation is \texttt{Sequence[PackageNode]}.}
\vspace{0.5em}
\label{fig:case-return-narrowing}
\end{subfigure}
\begin{subfigure}{\linewidth}
\begin{lstlisting}[style=pythoncode]
StyleType = Style | str

class Table:
  def __init__(self, border_style=""):
    ...

if __name__ == "__main__":
  table = Table(border_style="not dim")
\end{lstlisting}
\caption{Parameter type over-narrowing. The static analysis infers \texttt{str}, while the developer-written annotation is \texttt{StyleType}.}
\label{fig:case-param-narrowing}
\end{subfigure}
\caption{Examples where static analysis over-narrows developer-written types.}
\label{fig:case-weakness}
\vspace{-1em}
\end{figure}

\section{Discussion \& Threats to Validity}
\label{sec:discussion}

\subsection{Type-Checker Results}

\revise{B1, B4}{
\label{sec:type-checker-results}
Our evaluation treats predictions that differ from developer-written types as incorrect, even though alternative types can also be valid.
To complement this evaluation, we examine the alarms reported by \pyrefly{} for annotations generated by \tool{} and \aider{}.

\myparagraph{Type-Checker Alarms}
We count type-related alarms after excluding those already present in the original code.
Compared with \aider{}, \tool{} reduces the number of alarms from 2,049 to 1,865 for Gemma 4.
The count also decreases from 925 to 911 for gpt-oss, but increases slightly from 997 to 1,012 for GLM-4.7.
To examine these alarm counts in more detail, we also compare the use of standalone \texttt{Any}, which can suppress type-checker alarms.
Across the three models, \aider{} produces 82.3\% more standalone \texttt{Any} annotations than \tool{}.
Of these annotations, 66.7\% do not match the developer-written types for \aider{}, compared with 49.0\% for \tool{}.
In summary, \tool{} produces alarm counts comparable to those of \aider{}, at substantially lower \llm{} cost and with fewer standalone \texttt{Any} annotations.

\myparagraph{Analysis of Mismatches}
We also examine whether predictions that differ from developer-written types affect \pyrefly{} alarms.
For each mismatch, we replace the predicted type with \texttt{Any} to relax its type constraint and compare alarms before and after replacement.
A decrease in alarms indicates that the predicted type triggers type-checker errors.
Alarms remain unchanged for 70.7\% of \aider{} mismatches and 71.4\% of \tool{} mismatches.
These results suggest that \tool{} can offer plausible alternative type annotations at substantially lower \llm{} cost.

}

\subsection{Threats to Validity}
\label{sec:threats-validity}

\revise{C1, B4}{

\myparagraph{Construct Validity}
Our accuracy metric uses exact match, so it can penalize valid alternative types.
Although we complement this metric with type-checker analysis to assess annotation quality (Section~\ref{sec:type-checker-results}),
further evaluation is needed to assess developers' preferences for the inferred types.

\myparagraph{External Validity}
Our evaluation removes type annotations from 12 \typybench{} projects to provide a controlled setting for comparison.
However, this setting may not fully reflect naturally under-annotated repositories, and further evaluation on such repositories would help assess the broader applicability of our findings.
Since reference annotations are unavailable for missing types, such evaluation would rely on type-checker analysis or developer feedback on the inferred types.

\myparagraph{Internal Validity}
Static filling may insert incorrect types or propagate incorrect \llm{} predictions.
Approximately 2\% of slots receiving static type information are correctly annotated by \aider{} but not by \tool{}.
A post-processing step that uses type-checker feedback to correct such errors would help improve annotation quality.
In addition, \llm{} non-determinism may affect the reproducibility of results.
We release all \llm{} interaction logs to support reproducibility.

}
\section{Related Work}
\label{sec:related}

\myparagraph{Rule-based Type Inference}
Rule-based approaches infer types by applying hand-designed rules over program syntax and semantics~\cite{rule-alias,rule-maxsmt,rule-intepret,rule-ml,rule-tensor,rule-starkiller,rule-pyinder}.
This line of work has had substantial practical impact, especially in Python static analysis and type checking tools~\cite{pyright,mypy,pytype,pyrefly}.
However, Python programs often rely on dynamic features, implicit data flows, and project-specific conventions that are difficult to capture completely with fixed rules.

\myparagraph{Learning-based Type Inference}
Early learning-based systems predict types using supervised models, including graph neural networks and deep similarity models~\cite{lambdanet,deeplearning,class-dlinfer,class-typewriter,hityper,similar-type4py,similar-typlius}.
However, their predictions remain constrained by training distributions and type vocabularies, limiting their ability to infer complex annotations.
\revise{A1,B5}{
Recent work explores context construction for type inference with pre-trained code models and \llms{}~\cite{typet5,tiger,class-typebert,typegen,typepro}.
\typetfive{}~\cite{typet5} uses caller-callee context and incorporates previous type predictions through iterative decoding.
\typegen{}~\cite{typegen} combines dependency-based code slices, available types, and chain-of-thought examples in its prompts.
\textsc{TypePro}~\cite{typepro} uses static analysis to extract interprocedural code slices for its prompts.
A common challenge is to construct compact context while preserving relevant type information.
\tool{} supplements its code context with slot-specific hints that summarize type evidence from forward and backward propagation, reducing the need for supporting caller and callee code.
}


\myparagraph{Hybrid Approach to Type Inference}
Hybrid techniques combine learned models with static analysis to improve prediction quality~\cite{class-dlinfer,class-typewriter,hityper,nester,pytir}.
\textsc{DLInfer} \cite{class-dlinfer} and \textsc{Nester}~\cite{nester} extract dataflow graphs from the local context to provide additional information to the model.
\textsc{HiTyper}~\cite{hityper} applies forward type inference and type checking to validate model predictions.
\revise{A1, B5}{
To improve type correctness, \hityper{} selects type slots whose predictions enable forward inference for other unresolved slots.
This selection does not directly optimize query cost in \tool{}'s setting, where annotating a function can help resolve callers and callees through forward and backward propagation.
To reduce queries, \tool{} prioritizes functions with more distinct unresolved caller and callee neighbors.
It updates these priorities after each propagation step to reflect newly resolved functions.
\textsc{PyTIR}~\cite{pytir} iteratively refines types and dependencies across a repository, using type-checker feedback to re-query the \llm{} for conflicting annotations.
\tool{} reduces the query workload through direct static annotation and propagation.
The approaches are complementary and could be combined to reduce query costs while improving annotation consistency through type-checker feedback.
}

\section{Conclusion}
\label{sec:conclusion}

This paper presented \tool{}, a type annotation system that uses static analysis to guide \llm{} inference for \python{} programs.
Rather than invoking an \llm{} uniformly for every target function, 
\tool{} uses project-wide type evidence to resolve annotations statically, prioritize high-value \llm{} queries, and construct concise type-aware context.
Our evaluation on real-world \python{} projects shows that this combination
improves both accuracy and efficiency over specialized type inference systems
and strong coding-assistant baselines.
The results suggest that static analysis is a useful complement to \llm{}-based type inference: 
it preserves precise value-flow evidence, reduces unnecessary model calls, and provides concise context that improves the quality of \llm{} predictions.
This combination provides a practical way to improve type annotation for large \python{} codebases.

\section*{Data Availability}
For open science, we have made our implementation and evaluation data publicly available
on GitHub\footnote{\url{https://github.com/kupl/salt-artifact}}
and Zenodo\footnote{\url{https://doi.org/10.5281/zenodo.23788803}}.

\section*{Acknowledgment}
This work was supported by the German Research Foundation (DFG; projects 492507603, 516334526, and 526259073).
This work was partly supported by Institute of Information \& Communications Technology Planning
\& Evaluation (IITP) grant funded by the Korean government (MSIT) (No. 2020-0-01337, (SW STAR
LAB) Research on Highly-Practical Automated Software Repair)
and by ICT Creative Consilience Program through the Institute of Information \& Communications
Technology Planning \& Evaluation (IITP) grant funded by the Korean government (MSIT) (IITP-2026-RS-2020-II201819, 5\%).
This work was also supported by the National Research Foundation of
Korea (NRF) grant funded by the Korean government (MSIT) (No. 2021R1A5A1021944 (10\%), RS-2026-25470524 (50\%)).

\bibliographystyle{IEEEtran}
\bibliography{IEEEabrv, refs}

@article{wilcoxon,
 ISSN = {00994987},
 URL = {http://www.jstor.org/stable/3001968},
 author = {Frank Wilcoxon},
 journal = {Biometrics Bulletin},
 number = {6},
 pages = {80--83},
 publisher = {[International Biometric Society, Wiley]},
 title = {Individual Comparisons by Ranking Methods},
 urldate = {2026-10-22},
 volume = {1},
 year = {1945}
}

@inproceedings{tiger,
  title={Tiger: A generating-then-ranking framework for practical {Python} type inference},
  author={Wang, Chong and Zhang, Jian and Lou, Yiling and Liu, Mingwei and Sun, Weisong and Liu, Yang and Peng, Xin},
  booktitle={2025 IEEE/ACM 47th International Conference on Software Engineering (ICSE)},
  pages={321--333},
  year={2025},
  organization={IEEE}
}

@inproceedings{typet5,
  title={{TypeT5}: Seq2seq Type Inference using Static Analysis},
  author={Wei, Jiayi and Durrett, Greg and Dillig, Isil},
  booktitle={The Eleventh International Conference on Learning Representations},
  year={2023}
}

@inproceedings{hityper,
  title={Static inference meets deep learning: a hybrid type inference approach for {Python}},
  author={Peng, Yun and Gao, Cuiyun and Li, Zongjie and Gao, Bowei and Lo, David and Zhang, Qirun and Lyu, Michael},
  booktitle={Proceedings of the 44th International Conference on Software Engineering},
  pages={2019--2030},
  year={2022}
}

@inproceedings{typegen,
  title={Generative type inference for {Python}},
  author={Peng, Yun and Wang, Chaozheng and Wang, Wenxuan and Gao, Cuiyun and Lyu, Michael R},
  booktitle={2023 38th IEEE/ACM International Conference on Automated Software Engineering (ASE)},
  pages={988--999},
  year={2023},
  organization={IEEE}
}

@misc{aider,
  author = {{Aider-AI}},
  title = {Aider: {AI} Pair Programming in Your Terminal},
  howpublished = {\url{https://github.com/Aider-AI/aider}},
  note = {Accessed: 2026-06-26}
}

@misc{claudecode,
  author = {{Anthropics}},
  title = {{Claude Code}},
  howpublished = {\url{https://github.com/anthropics/claude-code}},
  note = {Accessed: 2026-06-26}
}

@misc{codexcli,
  author = {{OpenAI}},
  title = {{Codex CLI}},
  howpublished = {\url{https://github.com/openai/codex}},
  note = {Accessed: 2026-06-26}
}

@misc{githubcopilot,
  author = {{GitHub}},
  title = {{GitHub Copilot}},
  howpublished = {\url{https://github.com/features/copilot}},
  note = {Accessed: 2026-06-26}
}

@misc{cursor,
  author = {{Anysphere}},
  title = {Cursor},
  howpublished = {\url{https://cursor.com/}},
  note = {Accessed: 2026-06-26}
}

@inproceedings{ore2018assessing,
  title={Assessing the type annotation burden},
  author={Ore, John-Paul and Elbaum, Sebastian and Detweiler, Carrick and Karkazis, Lambros},
  booktitle={Proceedings of the 33rd ACM/IEEE International Conference on Automated Software Engineering},
  pages={190--201},
  year={2018}
}

@inproceedings{di2022evolution,
  title={The evolution of type annotations in {Python}: an empirical study},
  author={Di Grazia, Luca and Pradel, Michael},
  booktitle={Proceedings of the 30th ACM Joint European Software Engineering Conference and Symposium on the Foundations of Software Engineering},
  pages={209--220},
  year={2022}
}

@inproceedings{xie2025type,
  title={Do Type Annotations Help Students Improve {Python} Program Development Efficiency?},
  author={Xie, Wentian and Zhang, Sufan and Wang, Ziyuan},
  booktitle={2025 25th International Conference on Software Quality, Reliability, and Security Companion (QRS-C)},
  pages={594--600},
  year={2025},
  organization={IEEE}
}

@InProceedings{typybench,
  title = 	 {{T}ypy{B}ench: Evaluating {LLM} Type Inference for Untyped {Python} Repositories},
  author =       {Dong, Honghua and Yang, Jiacheng and Deng, Xun and Jiang, Yuhe and Pekhimenko, Gennady and Long, Fan and Si, Xujie},
  booktitle = 	 {Proceedings of the 42nd International Conference on Machine Learning},
  pages = 	 {14203--14230},
  year = 	 {2025},
  editor = 	 {Singh, Aarti and Fazel, Maryam and Hsu, Daniel and Lacoste-Julien, Simon and Berkenkamp, Felix and Maharaj, Tegan and Wagstaff, Kiri and Zhu, Jerry},
  volume = 	 {267},
  series = 	 {Proceedings of Machine Learning Research},
  month = 	 {13--19 Jul},
  publisher =    {PMLR},
  url = 	 {https://proceedings.mlr.press/v267/dong25l.html}
}

@misc{pyright,
  author = {Microsoft},
  title = {Pyright: Static type checker for {Python}},
  howpublished = {\url{https://github.com/microsoft/pyright}},
  note = {Accessed: 2026-06-28}
}

@misc{mypy,
  author = {Python},
  title = {mypy: Optional static typing for {Python}},
  howpublished = {\url{https://www.mypy-lang.org/}},
  note = {Accessed: 2026-06-28}
}

@misc{pytype,
  author = {Google},
  title = {pytype: A static type analyzer for {Python} code},
  howpublished = {\url{https://github.com/google/pytype}},
  note = {Accessed: 2026-06-28}
}

@misc{pyrefly,
  author = {Meta},
  title = {Pyrefly: A fast type checker and language server for {Python}},
  howpublished = {\url{https://github.com/facebook/pyrefly}},
  note = {Accessed: 2026-06-28}
}

@inproceedings{rule-alias,
  title = {Alias analysis for optimization of dynamic languages},
  author = {Gorbovitski, Michael and Liu, Yanhong A. and Stoller, Scott D. and Rothamel, Tom and Tekle, Tuncay K.},
  booktitle = {Proceedings of the 6th Symposium on Dynamic Languages},
  pages = {27--42},
  year = {2010}
}

@inproceedings{rule-maxsmt,
  title = {{MaxSMT}-based type inference for {Python} 3},
  author = {Hassan, Mostafa and Urban, Caterina and Eilers, Marco and M{\"u}ller, Peter},
  booktitle = {International Conference on Computer Aided Verification},
  pages = {12--19},
  year = {2018},
  organization = {Springer}
}

@inproceedings{rule-intepret,
  author = {Rigo, Armin and Pedroni, Samuele},
  title = {{PyPy}'s approach to virtual machine construction},
  booktitle = {Companion to the 21st ACM SIGPLAN Symposium on Object-Oriented Programming Systems, Languages, and Applications},
  pages = {944--953},
  year = {2006}
}

@inproceedings{rule-ml,
  author = {Dolby, Julian and Shinnar, Avraham and Allain, Allison and Reinen, Jenna},
  title = {Ariadne: analysis for machine learning programs},
  booktitle = {Proceedings of the 2nd ACM SIGPLAN International Workshop on Machine Learning and Programming Languages},
  pages = {1--10},
  year = {2018}
}

@inproceedings{rule-tensor,
  title = {Static analysis of shape in {TensorFlow} programs},
  author = {Lagouvardos, Sifis and Dolby, Julian and Grech, Neville and Antoniadis, Anastasios and Smaragdakis, Yannis},
  booktitle = {34th European Conference on Object-Oriented Programming},
  pages = {15:1--15:29},
  year = {2020}
}

@phdthesis{rule-starkiller,
  title = {Starkiller: A static type inferencer and compiler for {Python}},
  author = {Salib, Michael},
  school = {Massachusetts Institute of Technology},
  year = {2004}
}

@inproceedings{rule-pyinder,
  title = {Towards Effective Static Type-Error Detection for {Python}},
  author = {Oh, Wonseok and Oh, Hakjoo},
  booktitle = {Proceedings of the 39th IEEE/ACM International Conference on Automated Software Engineering},
  pages = {1808--1820},
  year = {2024}
}

@inproceedings{lambdanet,
  title = {{LambdaNet}: Probabilistic Type Inference using Graph Neural Networks},
  author = {Wei, Jiayi and Goyal, Maruth and Durrett, Greg and Dillig, Isil},
  booktitle = {International Conference on Learning Representations},
  year = {2020}
}

@inproceedings{deeplearning,
  title = {Deep learning type inference},
  author = {Hellendoorn, Vincent J. and Bird, Christian and Barr, Earl T. and Allamanis, Miltiadis},
  booktitle = {Proceedings of the 26th ACM Joint Meeting on European Software Engineering Conference and Symposium on the Foundations of Software Engineering},
  pages = {152--162},
  year = {2018}
}

@inproceedings{class-dlinfer,
  author = {Yan, Yanyan and Feng, Yang and Fan, Hongcheng and Xu, Baowen},
  title = {{DLInfer}: Deep Learning with Static Slicing for {Python} Type Inference},
  booktitle = {2023 IEEE/ACM 45th International Conference on Software Engineering},
  pages = {2009--2021},
  year = {2023}
}

@inproceedings{class-typewriter,
  author = {Pradel, Michael and Gousios, Georgios and Liu, Jason and Chandra, Satish},
  title = {{TypeWriter}: Neural Type Prediction with Search-based Validation},
  booktitle = {Proceedings of the 28th ACM Joint Meeting on European Software Engineering Conference and Symposium on the Foundations of Software Engineering},
  pages = {209--220},
  year = {2020}
}

@inproceedings{class-typebert,
  author = {Jesse, Kevin and Devanbu, Premkumar T. and Ahmed, Toufique},
  title = {Learning type annotation: is big data enough?},
  booktitle = {Proceedings of the 29th ACM Joint Meeting on European Software Engineering Conference and Symposium on the Foundations of Software Engineering},
  pages = {1483--1486},
  year = {2021}
}

@inproceedings{similar-type4py,
  author = {Mir, Amir M. and Lato\v{s}kinas, Evaldas and Proksch, Sebastian and Gousios, Georgios},
  title = {{Type4Py}: Practical Deep Similarity Learning-based Type Inference for {Python}},
  booktitle = {Proceedings of the 44th International Conference on Software Engineering},
  pages = {2241--2252},
  year = {2022}
}

@inproceedings{similar-typlius,
  author = {Allamanis, Miltiadis and Barr, Earl T. and Ducousso, Soline and Gao, Zheng},
  title = {{Typilus}: Neural Type Hints},
  booktitle = {Proceedings of the 41st ACM SIGPLAN Conference on Programming Language Design and Implementation},
  pages = {91--105},
  year = {2020}
}

@inproceedings{nester,
  title={Dataflow-Guided Neuro-Symbolic Language Models for Type Inference},
  author={Li, Gen and Wan, Yao and Zhang, Hongyu and Zhao, Zhou and Jiang, Wenbin and Shi, Xuanhua and Jin, Hai and Wang, Zheng},
  booktitle={International Conference on Machine Learning},
  pages={36000--36020},
  year={2025},
  organization={PMLR}
}

@misc{gemma4,
  title = {Gemma 4},
  author = {{Google DeepMind}},
  howpublished = {\url{https://deepmind.google/models/gemma/gemma-4/}},
  note = {Accessed: 2026-06-24}
}

@misc{gptoss,
  title = {{gpt-oss}},
  author = {{OpenAI}},
  howpublished = {\url{https://openai.com/index/introducing-gpt-oss/}},
  note = {Accessed: 2026-06-24}
}

@misc{glm47,
  title = {{GLM-4.7}},
  author = {{Z.ai}},
  howpublished = {\url{https://z.ai/blog/glm-4.7}},
  note = {Accessed: 2026-06-24}
}

@misc{jedi,
  title = {Jedi: An awesome autocompletion, static analysis and refactoring library for {Python}},
  author = {{Dave Halter}},
  howpublished = {\url{https://github.com/davidhalter/jedi}},
  note = {Accessed: 2026-06-25}
}

@article{pytir,
author = {Sun, Shuo and Zhang, Shixin and Yan, Jiwei and Yan, Jun and Zhang, Jian},
title = {Co-evolution of Types and Dependencies: Towards Repository-Level Type Inference for {Python} Code},
year = {2026},
issue_date = {July 2026},
publisher = {Association for Computing Machinery},
address = {New York, NY, USA},
volume = {3},
number = {FSE},
url = {https://doi.org/10.1145/3797103},
doi = {10.1145/3797103},
journal = {Proc. ACM Softw. Eng.},
month = jun,
articleno = {FSE075},
numpages = {21}
}

@article{typepro,
author = {Lin, Teyu and Fan, Minghao and Huang, Huaxun and Shen, Zhirong and Wu, Rongxin},
title = {{TypePro}: Boosting {LLM}-Based Type Inference via Inter-Procedural Slicing},
year = {2026},
issue_date = {July 2026},
publisher = {Association for Computing Machinery},
address = {New York, NY, USA},
volume = {3},
number = {FSE},
url = {https://doi.org/10.1145/3808124},
doi = {10.1145/3808124},
journal = {Proc. ACM Softw. Eng.},
month = jun,
articleno = {FSE117},
numpages = {22}
}

\end{document}